\documentclass[iop, apj]{emulateapj}
\shorttitle{Hard X-ray PSD's of AGN}
\shortauthors{Shimizu, Mushotzky}
\submitted{Accepted for publication in ApJ on April 23, 2013}
\usepackage[colorlinks,urlcolor=blue,citecolor=blue,linkcolor=blue]{hyperref}
\usepackage{color}
\begin{document}

\title{The First Hard X-Ray Power Spectral Density Functions of AGN}
\author{T. Taro Shimizu, Richard F. Mushotzky}
\affil{Department of Astronomy, University of Maryland, College Park, MD 20742, USA}
\email{tshimizu@astro.umd.edu}

\begin{abstract}
We present results of our Power Spectral Density (PSD) analysis of 30 AGN using the 58 month light curves from \textit{Swift's} Burst Alert Telescope (BAT) in the 14--150 keV band. PSDs were fit using a Monte Carlo based algorithm to take into account windowing effects and measurement error. All but one source were found to be fit very well using an unbroken power law with a slope of $\sim-1$, consistent at low frequencies with previous studies in the 2--10 keV band, with no evidence of a break in the PSD. For 5 of the highest S/N sources we tested the energy dependence of the PSD and found no significant difference in the PSD at different energies. Unlike previous studies of X-ray variability in AGN, we do not find any significant correlations between the hard X-ray variability and different properties of the AGN including luminosity and black hole mass. The lack of break frequencies and correlations seem to indicate that AGN are similar to the high state of Galactic Black Holes.
\end{abstract}


\section{Introduction}\label{sect_1}
Active Galactic Nuclei (AGN) are thought to be powered by accretion onto a Supermassive Black Hole (SMBH) because of their high luminosity and small size. The gravitational potential energy of infalling material in the accretion disk is converted into thermal energy and radiated away. However infall only explains AGN spectra up to UV energies whereas AGN are also seen to emit strongly in X-rays. To explain this, a corona must be invoked that is thought to Compton upscatter the UV photons from the inner accretion disk to X-ray energies. The upscattered photons are radiated quasi-isotropically so some directly travel to the observer or illuminate the accretion disk. This illumination produces the so-called "Compton hump" seen in the spectra of AGN that peaks around 30 keV (eg. \citet{1991MNRAS.249..352G, 1994MNRAS.268..405N} for a theoretical discussion and \citet{1995MNRAS.273..837M, 2011ApJS..193....3R} for observational evidence.)

AGN X-ray emission also exhibits very strong and rapid variability that suggests the origin of the emission is very near the SMBH itself. By studying the spectral and temporal properties of the X-rays, we can learn about the environment inside the strongest gravitational potential wells in the universe.

Variability has been a known property of AGN for several decades. \citet{1986Natur.320..421B} found that for Seyfert galaxies the timescale of X-ray variability is inversely related to the X-ray luminosity. Using EXOSAT, \citet{1987Natur.325..694L,1993ApJ...414L..85L,1993MNRAS.265..664G} were the first to be able to characterize the variability as aperiodic following a power law relationship with frequency, $P(\nu) \propto \nu^{-\alpha}$,  with $P(\nu)$ being the power at a given frequency and $\alpha\sim2$. 

The most powerful and often used tool in studying time variability is the Power Spectral Density function (PSD), which measures the contribution to the total variance that each temporal frequency provides. \citet{1995MNRAS.273..923P} created the first broadband PSD with EXOSAT observations while \citet{1999ApJ...514..682E}, using the new {\textit Rossi X-ray Timing Explorer (RXTE)} that provided well sampled X-ray light curves, constructed the first high quality broadband PSD and measured the first ``break'' in the PSD where $\alpha$ changed from a value of 1.74 to 0.73 as temporal frequency decreases. The slopes measured were later found to be consistent with 2 and 1 respectively due to \citet{1999ApJ...514..682E} not accounting for aliasing and a miscalculation of errors on the PSD \citep{2002MNRAS.332..231U}. This break was theoretically needed or else the variability would diverge as timescales increased.

Since \citet{1999ApJ...514..682E} there have been numerous PSD studies of AGN \citep{2002MNRAS.332..231U,2005MNRAS.363..586U,2004MNRAS.348..783M,2005MNRAS.359.1469M,2003ApJ...593...96M} all aimed at studying the long and short term variability of AGN, measuring the break frequencies and power law slopes of the PSD. From all these studies, a relationship emerged between the break frequency and black hole mass, however not only for AGN but including Galactic Black Holes (GBH) \citep{2004MNRAS.348..207P,2005MNRAS.364..208D} . Since AGN and GBH both harbor black holes at their centers, it had long been postulated that AGN are just scaled up versions of the same system with the same processes governing both. Break frequencies also occur in the PSDs of GBHs and using simple scaling with black hole mass, a rough correlation was found leading to speculation the break frequency is associated with the inner edge of the accretion disk. However there was a large scatter in the correlation indicating that a second variable was needed to strengthen relationship. In particular it was seen that at the same black hole mass, higher accretion rate Narrow Line Seyfert 1's (NLS1) displayed a higher break frequency than ``normal'' Seyfert 1s, so accretion rate (actually bolometric luminosity to make all parameters independent) was introduced into the relationship. These two properties of black hole mass and accretion rate correctly account for the scaling of break frequency from stellar mass X-ray binaries up to supermassive AGN. \citep{2006Natur.444..730M}. 

All of these PSD studies, however focus on the 2-10 keV energy range because that was the best band for {\it RXTE}. This band though can be affected by absorption, so most of the samples have been limited to Seyfert 1's since these, according to the unified model, have a clear view of the accretion disk. Harder energies with E $>$ 10 keV, though, ``see through'' any gas as long as it is ``Compton thin'' with hydrogen column density, $N_{\rm{H}} < ~3\times10^{24}\, \,\rm{cm^{-2}}$. With the {\it Swift} satellite and its Burst Alert Telescope (BAT), we can now begin to study the hard X-ray sky in an unbiased way because its energy range is 14--195 keV. BAT has been continuously monitoring the sky with a wide field of view for $\sim$6 years, thus providing high quality light curves for many sources for the use of PSD analysis.

Since Compton reflection peaks in this energy range we can also begin to answer the question of whether the continuum and reflection components of AGN spectra show different variability characteristics. \citet{2009MNRAS.399.1597S} analyzed {\it RXTE} spectra of 10 AGN and found that the model that best described the variability was one with a constant reflection component and an intrinsic power law that varies in both flux and power law index. \citet{2007PASJ...59S.315M} studied the Seyfert 1 galaxy, MCG-6-30-15 and also found reduced variability in the 14--45 keV band where reflection peaks in the energy spectrum. 

\citet{2007A&A...475..827B} looked at the hard X-ray variability using the 9 month BAT catalog, however the 9 month light curves weren't of high enough quality for a PSD study. Instead two other tools, excess variance and structure function were used in their analysis. They found that absorbed, Seyfert 2 galaxies show more variability than unabsorbed, Seyfert 1 galaxies, which is surprising since unless the AGN is Compton thick there should not be a large effect on the variability especially at these long timescales. They also reported an anticorrelation between luminosity and variability, similar to that seen in softer X-rays suggesting the same variability process is working in the 2--10 keV energies as in 14--195 keV energies. However recent work by \citet{2010MNRAS.404..931E} have highlighted the unreliability of structure functions in determining variability properties.

With the 58 month light curves now available, we construct and measure the PSDs of a sample of BAT selected AGN to further study the hard X-ray variability. We provide the very first PSD slopes and normalizations in the 14-150 keV energy band and compare the PSD measured variability with AGN properties to determine correlations. Because of the large energy range of BAT, we also are able to test any underlying energy dependence of the PSD as suggested by \citet{2007ApJ...656..116M} for Mrk 766 and \citet{1997ApJ...474L..57C}for Cygnus X-1.

The paper is organized as follows. \S2 details the BAT observations and reduction of light curves including rebinning and filtering. \S3 describes the construction of PSDs, selection of sample, and determination of PSD parameters. \S4 highlights the results of the model fitting including the energy dependence of the PSD and correlations between AGN properties and variance. \S5 discusses the implications of these results and compares them to previous studies of AGN variability. 

\section{Data Reduction}\label{sect_2}
Since November 2004, {\it Swift} \citep{2004ApJ...611.1005G} with the Burst Alert Telescope (BAT) \citep{2005SSRv..120..143B} has been searching for gamma-ray bursts (GRBs) in the hard X-ray sky (14--195 keV). {\it Swift}/BAT uses a 5200 cm$^{2}$ coded-aperture mask above an array of 32768 CdZnTe detectors to produce a wide field of view ($~\sim1$ str fully coded, 3 sr partially coded) of the sky. When a GRB appears, BAT software can quickly locate the position to within 1-4 arcmin using a transformation algorithm and allow other instruments and telescopes to observe the GRB. 

However while in survey mode (ie not specifically observing a GRB), BAT is continuously observing the sky. Every five minutes a new snapshot image is produced, and these series of snapshot images are then cleaned so count rates can be extracted for sources that were within the field of view. In this way, snapshot images are turned into light curves for a large number of hard X-ray sources due to the wide field of view and large sky coverage of BAT. (See \citet{2010ApJS..186..378T} for details of cleaning process.) Each lightcurve contains count rates over a ${\sim}$5 year period in eight different energy bands: 14--20, 20--24, 24--35, 35--50, 50--75, 75--100, 100--150, 150--195 keV, and then a total count rate for the entire 14--195 keV band with each datapoint corresponding to single 5 minute snapshot image.

\begin{figure*}[t!]
\centering
\includegraphics[width = \textwidth]{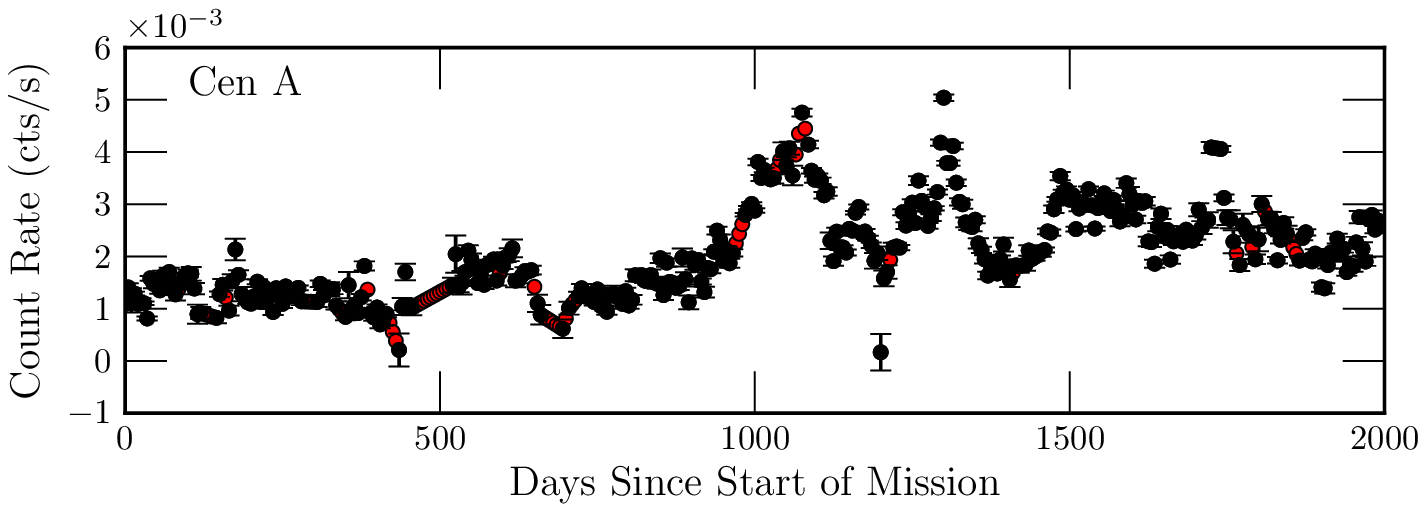}
\includegraphics[width = \textwidth]{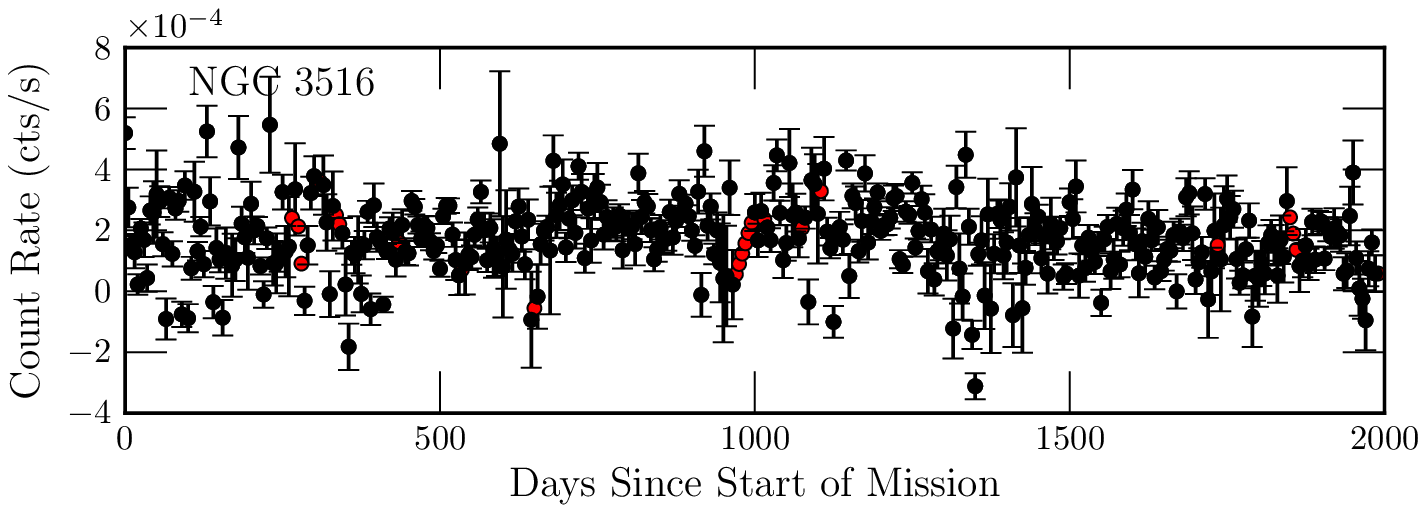}
\caption{{\it Swift}/BAT light curves of Centaurus A and NGC 3516 after rebinning, filtering, interpolation. Interpolated points are shown in red without error bars. \label{fig_0}}
\end{figure*}

For our PSD analysis, we first downloaded the 58-month raw snapshot lightcurves for each AGN provided by \citet{baum11} on HEASARC\footnote{\url{http://heasarc.nasa.gov/docs/swift/results/bs58mon/}}. We then added together the count rates of the first 7 bands, producing a 14--150 keV band light curve to maximize S/N in the PSD. We decided to discard the 150--195 keV band data, because we felt it did not add any information to the variability of the sources due to the low S/N of the light curve at these high energies.\footnote{Private communication with Jack Tueller and Wayne Baumgartner} The raw lightcurves were rebinned into time bins with a width of 5 days using a weighted average so that the PSD was above white noise level that overwhelms at high temporal frequencies due to Poisson noise. With this rebinning, our PSDs span the frequency range $10^{-8} - 10^{-5.9}$ Hz, 2 orders of magnitude. Also any bin in the light curve with a fractional exposure $<2\%$ was not used in the analysis to ensure Gaussian statistics were still a good assumption. A cut of $<2\%$ was found to routinely keep the number of raw light curve points in each bin above 20, but still allowed for some bins to be kept that had less than 20 but a large exposure time. Gaps in the light curves due to the filtering process, South Atlantic Anomaly, GRB observations, and satellite maintenance, were filled by linearly interpolating between the edges of the gaps without significantly altering the PSD. If too many bins are missing in the light curve, linear interpolation could cause an artificial smoothing and reduce the variability at high frequencies. All of the sources in our analysis had less than $\sim$20$\%$ of their light curve missing due to gaps. Figure~\ref{fig_0} shows the light curves of a high S/N source (Centaurus A) and a source 10 times fainter (NGC 3516) after all rebinning, filtering, and interpolation have taken place. Notice there is still distinct variability even at long timescales. 

\section{Data Analysis}\label{sect_3}
\subsection{Power Spectral Density Function}\label{subsect_3.1}
\begin{figure}[t!]
\epsscale{1.0}
\plotone{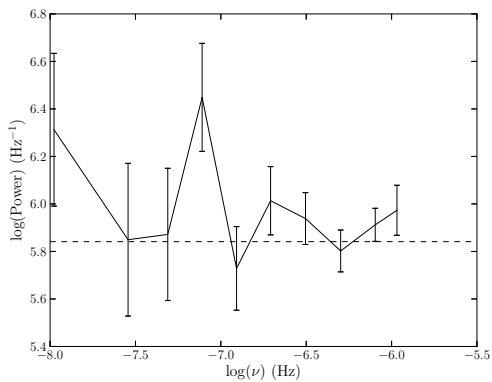}
\caption{Observed PSD of the Coma galaxy cluster (solid line with error bars) plotted with the estimation of the Poisson noise contribution using Equation~\ref{eq_3}.\label{fig_1}}
\end{figure}

To construct the PSD, we first subtracted the mean count rate, $\mu$, from the light curve, which has the effect of removing the zero frequency power from the PSD. We then estimate the PSD using the periodogram which is the squared modulus of the Discrete Fourier Transform \citep{1975dsp..book.....O},
\begin{equation}\label{eq_1}
\left|F(\nu_j)\right|^2 = \left[\sum_{i=1}^{N}x_i\cos(2\pi\nu_jt_i)\right]^2 + \left[\sum_{i=1}^{N}x_i\sin(2\pi\nu_jt_i)\right]^2
\end{equation}
where $\nu_j = \frac{j}{T},j=1\ldots\frac{N}{2T}$, $T$ is total duration of the light curve, $N$ is the number of bins in the light curve, and $x_i$ is the count rate at time $t_i$. The periodogram is then defined as
\begin{equation}\label{eq_2}
P(\nu_j) = \frac{2T}{\mu^2N^2}\left|F(\nu_j)\right|^2
\end{equation}
With this normalization, $\frac{2T}{\mu^2N^2}$, known as fractional rms-squared normalization, periodograms from different segments of the light curve can be equally compared as well as PSDs from different AGN. Also the integral of the periodogram over a frequency range is equal to the contribution to the fractional rms-squared variance due to variations from the corresponding timescales \citep{1991ApJ...383..784M,1997scma.conf..321V}. However variability is also introduced due to measurement errors and will manifest itself as white noise (equal amplitude at all frequencies) in the PSD and cause flattening of the observed PSD at high frequencies. The expected noise level in the PSD from a light curve with Gaussian errors is given by \citep{2003MNRAS.345.1271V}.
\begin{equation}\label{eq_3}
P_{\rm{noise}} = \frac{2\Delta T \overline{\sigma^2_{\rm{err}}}}{\mu^2}
\end{equation}
$\overline{\sigma^2_{\rm{err}}} = \sum_i\sigma_{\rm{err},i}^2$ where $\sigma_{\rm{err},i}$ is the error on the count rate at time $t_i$. To test this estimate of the noise level, we constructed PSDs of a few clusters of galaxies. Clusters are constant sources of hard X-ray emission \citep{Wik:2011fr} so the PSD should be fairly flat at the $P_{noise}$ level. Figure~\ref{fig_1} shows the PSD of the Coma cluster, one of the brightest galaxy clusters in the BAT catalog. As can be seen, Coma's PSD (solid line with error bars) is approximately flat at the noise level (dashed line)  within the error bars that were calculated using the method from  \citet{1993MNRAS.261..612P}. Therefore Equation~\ref{eq_3} should be a good estimate of the white noise variability power from measurement errors. 

Following \citet{1993MNRAS.261..612P}, we then logarithmically binned the periodogram every factor of 1.6 in frequency creating equally spaced bins of width $\sim$0.2 in logarithm. The lowest two frequency bins however were widened so there were at least 3 original raw PSD points used in the average. As \citet{1993MNRAS.261..612P} also show, a bias of $\sim$-.253 is introduced into the power of the periodogram so this is corrected for at this stage.

To determine which sources to use in our model fitting, we constructed the observed PSDs of all the AGN in the BAT catalog that had a S/N greater than 20 (76 sources) using Equation~\ref{eq_2} and used visual inspection to look for a significant detection of power in the PSD well above $P_{\rm{noise}}$.  S/N was taken from the BAT catalog and is the significance of source detection in the BAT sky map. 30 sources showed a significant detection of power and thus qualified for further analysis with PSD fitting and are listed in Table~\ref{tbl_1} along with the properties of the BAT observation over the length of the mission. All signal -to-noise ratios and AGN classifications were taken from the BAT 58 month catalog\footnote{\url{http://heasarc.nasa.gov/docs/swift/results/bs58mon/}}. Mean count rates were determined by arithmetically averaging all the count rates together after rebinning and filtering. Missing fraction was determined by taking the ratio of the number of time bins in the rebinned light curve to the number of bins expected if the light curve was uniformly and fully sampled. 

\begin{deluxetable*}{llccc}]htbp]
\tabletypesize{\scriptsize}
\tablecaption{AGN Observation Properties\label{tbl_1}}
\tablewidth{0pt}
\tablehead{\colhead{} & \colhead{} & \colhead{} & \colhead{Mean Count Rate} & \colhead{Fraction} \\
\colhead{Name} & \colhead{Type} & \colhead{S/N} & \colhead{(cts s$^{-1}$)} & \colhead{Missing} \\
\colhead{(1)} & \colhead{(2)} & \colhead{(3)} & \colhead{(4)} & \colhead{(5)}}
\startdata
3C 273  & Blazar  & 156.8 & 6.38E-04 & 0.17 \\
3C 390.3  & Sy 1  & 47.97 & 1.64E-04 & 0.07 \\
3C 454.3  & Blazar  & 40.43 & 1.83E-04 & 0.15 \\
4U 1344-60  & Sy 1.5  & 38.73 & 1.66E-04 & 0.12 \\
Cen A  & NLRG  & 428.7 & 2.08E-03 & 0.16 \\
Circinus Galaxy  & Sy 2  & 101.7 & 4.62E-04 & 0.10 \\
Cygnus A  & NLRG  & 54    & 2.14E-04 & 0.08 \\
ESO 506-G027  & Sy 2  & 26.44 & 1.38E-04 & 0.17 \\
IC 4329A  & Sy 1  & 101.1 & 5.03E-04 & 0.18 \\
IGR J21247+5058  & BLRG  & 83.3  & 3.27E-04 & 0.08 \\
MCG-05-23-016  & Sy 1.9  & 90.4  & 3.55E-04 & 0.13 \\
MCG+08-11-011  & Sy 1.5  & 49    & 2.31E-04 & 0.19 \\
Mrk 110  & Sy 1  & 30.42 & 9.49E-05 & 0.10 \\
Mrk 3  & Sy 2  & 48.5  & 2.12E-04 & 0.07 \\
Mrk 348  & Sy 2  & 70.4  & 2.58E-04 & 0.13 \\
Mrk 421  & Blazar  & 109.5 & 3.10E-04 & 0.10 \\
Mrk 6  & Sy 1.5  & 26.75 & 9.02E-05 & 0.07 \\
Mrk 926  & Sy 1.5  & 45.57 & 1.80E-04 & 0.19 \\
NGC 2110  & Sy 2  & 98.1  & 4.98E-04 & 0.13 \\
NGC 3227  & Sy 1.5  & 56.2  & 1.73E-04 & 0.14 \\
NGC 3516  & Sy 1.5  & 62.3  & 1.76E-04 & 0.06 \\
NGC 3783  & Sy 1.5  & 68.7  & 2.66E-04 & 0.16 \\
NGC 4151  & Sy 1.5  & 275   & 8.37E-04 & 0.07 \\
NGC 4388  & Sy 2  & 110.7 & 4.22E-04 & 0.12 \\
NGC 4945  & Sy 2  & 76.1  & 3.72E-04 & 0.13 \\
NGC 5252  & Sy 2  & 42.4  & 1.81E-04 & 0.15 \\
NGC 5548  & Sy 1.5  & 32.55 & 1.27E-04 & 0.07 \\
NGC 6814  & Sy 2  & 22.52 & 1.14E-04 & 0.18 \\
NGC 4507  & Sy 2  & 64.6  & 2.82E-04 & 0.16 \\
NGC 7172  & Sy 2  & 60.1  & 2.48E-04 & 0.21 \\
\enddata
\tablecomments{Col.(1): Name of AGN. Col.(2): Type of AGN. Col.(3): Signal-to-noise ratio. Col.(4): Mean count rate over the whole light curve. Col.(5): Fraction of missing data due to gaps in the light curve.}
\end{deluxetable*}

\begin{deluxetable*}{llcccccccc}[t!]
\tabletypesize{\scriptsize}
\tablecaption{AGN Source Properties\label{tbl_2}}
\tablewidth{0pt}
\tablehead{\colhead{} & \colhead{} & \colhead{$\log L_{\rm{X}}$} & \colhead{$\log L_{\rm{bol}}$} & \colhead{$\log M_{\rm{BH}}$} & \colhead{} & \colhead{} & \colhead{} & \colhead{$\log N_{\rm{H}}$} & \colhead{}\\
\colhead{Name} & \colhead{Type} & \colhead{($\rm{ergs \,s}^{-1}$)} & \colhead{($\rm{ergs \,s}^{-1}$)} & \colhead{($M_{\sun}$)} & \colhead{Reference} & \colhead{Method} & \colhead{$L_{\rm{bol}}/L_{\rm{edd}}$} & \colhead{(cm$^{-2}$)} & \colhead{Reference}\\
\colhead{(1)} & \colhead{(2)} & \colhead{(3)} & \colhead{(4)} & \colhead{(5)} & \colhead{(6)} & \colhead{(7)} & \colhead{(8)} & \colhead{(9)} & \colhead{(10)}}
\startdata
    3C 273 & Blazar & 46.47 & 47.62 & 8.95$\pm$0.09  & P04   & RM    & 3.61  & 20.5  & B07 \\
    3C 390.3 & Sy 1  & 44.89 & 45.86 & 8.44$\pm$0.14  & W10   & RM    & 0.20  & 21.1  & Wi09 \\
    3C 454.3 & Blazar & 47.67 & 48.96 & 9.18$\pm$0.30  & W02   & R-L   & 46.08 & 20.9  & Wi09 \\
    4U 1344-60 & Sy 1.5 & 43.6  & 44.42 & 7.44$\pm$0.10  & V10   & M-L   & 0.07  & 22.2  & Wi09 \\
    Cen A & NLRG  & 43.01 & 43.76 & 7.85$\pm$0.24  & C09, N07, K07 & Stars & 0.01  & 22.7  & B11 \\
    Circinus Galaxy & Sy 2  & 42.09 & 42.73 & 6.23$\pm$0.08  & G03   & Maser & 0.02  & 24.6  & A12 \\
    Cygnus A & NLRG  & 45.01 & 45.99 & 9.40$\pm$0.12  & T03   & Gas   & 0.03  & 23.0  & Wi09 \\
    ESO 506-G027 & Sy 2  & 44.12 & 45.00 & 8.02$\pm$0.10  & V10   & M-L   & 0.07  & 23.9  & Wi09 \\
    IC 4329A & Sy 1  & 44.22 & 45.11 & 8.34$\pm$0.21  & O99, T02 & M-$\sigma$ & 0.05  & 21.8  & Wi09 \\
    IGR J21247+5058 & BLRG  & 44.25 & 45.14 & 6.58$\pm$0.07  & Wi10  & R-L   & 2.80  & 22.1  & Wi09 \\
    MCG-05-23-016 & Sy 1.9 & 43.5  & 44.30 & 6.29$\pm$0.70  & WZ07  & Gas   & 0.80  & 22.2  & Wi09 \\
    MCG+08-11-011 & Sy 1.5 & 44.09 & 44.96 & 8.07$\pm$0.02  & Wi10  & R-L   & 0.06  & 21.4  & Wi09 \\
    Mrk 110 & Sy 1  & 44.21 & 45.10 & 7.38$\pm$0.14  & W10   & RM    & 0.40  & 20.3  & Wi09 \\
    Mrk 3 & Sy 2  & 43.74 & 44.57 & 7.72$\pm$0.10  & V10   & M-L   & 0.05  & 22.1  & Wi09 \\
    Mrk 348 & Sy 2  & 43.91 & 44.76 & 7.41$\pm$0.10  & V10   & M-L   & 0.17  & 23.2  & Wi09 \\
    Mrk 421 & Blazar & 44.46 & 45.38 & 8.56$\pm$0.18  & W08   & M-$\sigma$ & 0.05  & 20.9  & Wi09 \\
    Mrk 6 & Sy 1.5 & 43.69 & 44.52 & 8.09$\pm$0.02  & Wi10  & R-L   & 0.02  & 22.5  & Wi09 \\
    Mrk 926 & Sy 1.5 & 44.76 & 45.71 & 8.36$\pm$0.02  & Wi10  & R-L   & 0.17  & 21.1  & B11 \\
    NGC 2110 & Sy 2  & 43.6  & 44.42 & 7.40$\pm$0.10  & V10   & R-L   & 0.08  & 22.5  & Wi09 \\
    NGC 3227 & Sy 1.5 & 42.57 & 43.27 & 7.60$\pm$0.24  & W10   & RM    & 0.00  & 22.2  & Wi09 \\
    NGC 3516 & Sy 1.5 & 43.33 & 44.12 & 7.61$\pm$0.18  & W10   & RM    & 0.02  & 21.5  & Wi09 \\
    NGC 3783 & Sy 1.5 & 43.6  & 44.42 & 7.45$\pm$0.13  & W10   & RM    & 0.07  & 21.8  & Wi09 \\
    NGC 4151 & Sy 1.5 & 43.11 & 43.87 & 7.64$\pm$0.11  & W10   & RM    & 0.01  & 22.7  & Wi09 \\
    NGC 4388 & Sy 2  & 43.64 & 44.46 & 7.07$\pm$0.10  & V10   & M-L   & 0.19  & 23.6  & Wi09 \\
    NGC 4945 & Sy 2  & 42.37 & 43.04 & 6.15$\pm$ 0.16  & G97   & Maser & 0.06  & 24.3  & A12 \\
    NGC 5252 & Sy 2  & 44.12 & 45.00 & 8.98$\pm$0.21  & C05   & Gas   & 0.01  & 22.6  & Wi09 \\
    NGC 5548 & Sy 1.5 & 43.69 & 44.52 & 7.89$\pm$0.10  & W10   & RM    & 0.03  & 20.8  & Wi09 \\
    NGC 6814 & Sy 2  & 43.24 & 44.01 & 7.25$\pm$0.12  & W10   & RM    & 0.04  & 20.8  & B11 \\
    NGC 4507 & Sy 2  & 43.77 & 44.61 & 7.70$\pm$0.10  & V10   & M-L   & 0.06  & 23.5  & Wi09 \\
    NGC 7172 & Sy 2  & 43.46 & 44.26 & 7.36$\pm$0.10  & V10   & M-L   & 0.06  & 22.9  & Wi09 \\
 \enddata
\tablecomments{Col.(1): Source name. Col.(2): AGN type taken from HEASARC. Col.(3): 14-195 keV x-ray luminosity from HEASARC. Col.(4): Bolometric luminosity derived using Equation~\ref{eq_2}. Col.(5): Black hole mass. Col.(6): Reference for black hole mass. Col.(7): Method used to determine black hole mass. RM = reverberation mapping, R-L = BLR kinematics using the $R_{\rm{BLR}}$ - optical luminosity relation, M-L = mass - bulge luminosity relation, Stars = stellar dynamics, Maser = maser dynamics, Gas = gas dynamics, M-$\sigma$ = mass - stellar velocity dispersion relation. Col.(8): Ratio of bolometric luminosity to Eddington luminosity. Eddington luminosity is determined using Equation~\ref{eq_3}. Col.(9): Column density. Col.(10): Reference for column density.}
\tablerefs{P04: \citet{2004ApJ...613..682P}; W10: \citet{2010ApJ...716..269W}; W02: \citet{2002ApJ...579..530W}; V10: \citet{2010MNRAS.402.1081V}; C09: \citet{2009MNRAS.394..660C}; N07: \citet{2007ApJ...671.1329N}; K07: \citet{2007MNRAS.374..385K}; G03: \citet{2003ApJ...590..162G}; T03: \citet{2003MNRAS.342..861T}; O99: \citet{1999A&A...350....9O}; T02: \citet{2002ApJ...574..740T}; WZ07: \citet{2007ApJ...660.1072W}; W08: \citet{2008MNRAS.385..119W}; G97: \citet{1997ApJ...481L..23G}; C05: \citet{2005A&A...431..465C}; B07: \citet{2007A&A...475..827B}; Wi09: \citet{2009ApJ...690.1322W}; B11: \citet{2011ApJ...728...58B}; A12: \citet{2012ApJ...749...21A}}
\end{deluxetable*}

\begin{figure}[t!]
\epsscale{1.0}
\plotone{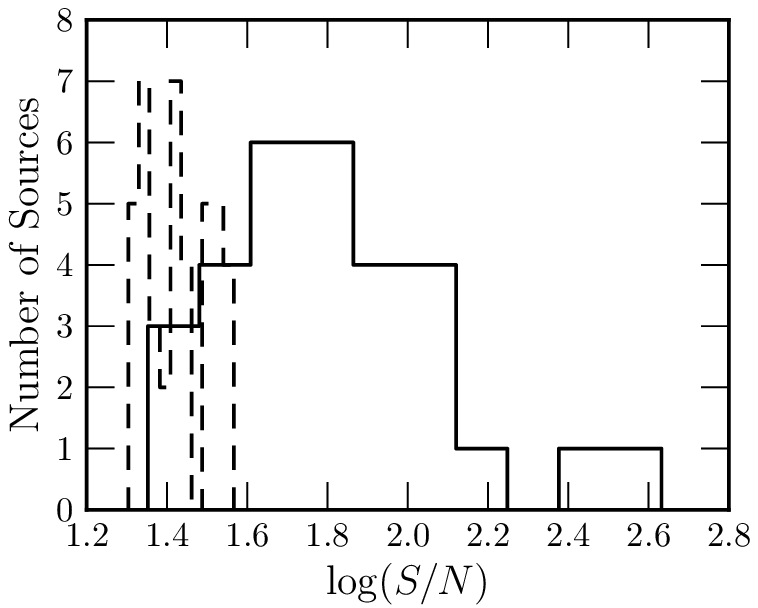}
\plotone{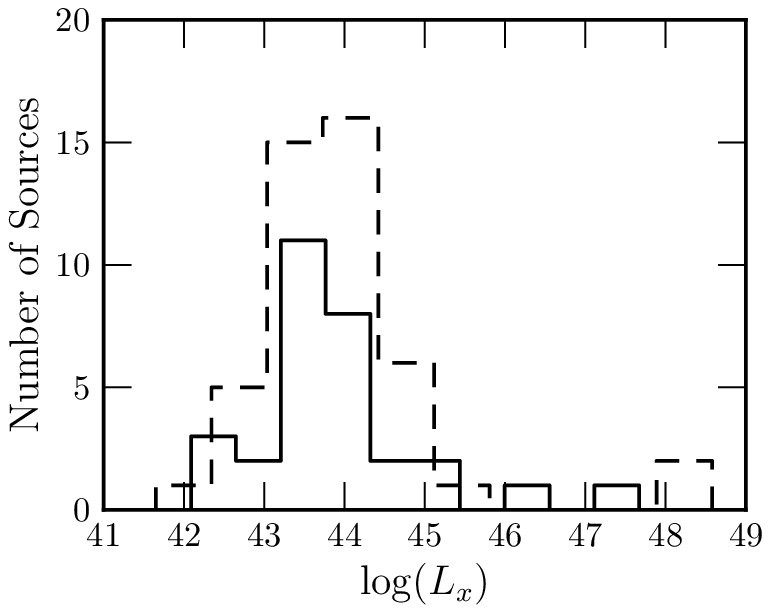}
\caption{Top: Distribution of S/N for the sample selected for PSD fitting (solid line) and the sample not selected (dashed line). Bottom: Distribution of luminosity for the sample selected for PSD fitting (solid line) and the sample not selected (dashed line). \label{fig_A}}
\end{figure}

In total our sample consists of 3 Seyfert 1's, 11 Seyfert 2's, 9 Seyfert 1.5's, 1 Seyfert 1.9, 3 Blazars, 2 NLRG's, and 1 BLRG. Because our sample was selected through visual inspection, there is an inherent bias towards bright and/or highly variable objects. Blazars are known to be highly variable, much more than Seyferts so these were expected. But what is interesting is many more Seyfert 2's than Seyfert 1's were selected. Our original 76 source sample from the catalog contained 33\% Seyfert 2's (25/76), 28\% Seyfert 1's (21/76), and 20\% Seyfert 1.5's whereas the sample we are fitting has 36\% Seyfert 2's, 10\% Seyfert 1's, and 30\% Seyfert 1.5's. However it seems that the sample we have selected is mainly those sources that have a high S/N so the distribution difference is more of an indication that there are more high S/N Seyfert 2's than there are Seyfert 1's. Figure~\ref{fig_A} shows the distribution of S/N for our selected sample and those that were not selected. We clearly have selected those objects that have a high S/N because these objects are the ones that will most likely have enough intrinsic variability above that of noise assuming that there is not a radical difference in intrinsic source variance. Figure~\ref{fig_A} also shows the luminosity distribution between the selected and unselected sample. It seems we are not biased towards high or low luminosity which means if there exists a correlation between the variance and luminosity we should be able to observe it.

Table~\ref{tbl_2} contains properties of each AGN determined from previous studies. X-ray luminosity was taken from HEASARC. Bolometric luminosity was calculated using the strong correlation between $L_{\rm{bol}}$ and $L_{\rm{X}}$ empirically found by \citet{2012ApJ...745..107W}.
\begin{equation}\label{eq_4}
\log(L_{\rm bol})=1.1157\log(L_{\rm X}) - 4.228
\end{equation}
This relation however was only found for Seyfert 1-1.5 so it is possible that for Seyfert 2's this is an underestimate of the bolometric luminosity since they have a higher hydrogen column density which absorbs the X-ray emission. However for the BAT band, $N_{\rm H}$ needs to be $>10^{24}$ cm$^{2}$ to have a significant effect. In our sample only 2 sources, NGC 4945 and Mrk 3, have a column density this high so we don't expect a serious discrepancy between the true bolometric luminosity and our estimate.

Black hole masses were all taken from the literature as well as $N_{\rm{H}}$, the intrinsic neutral hydrogen column density. The references for both $M_{\rm{BH}}$ and $N_{\rm{H}}$ are listed in the table along with the method used to determine each black hole mass. For $L_{\rm{edd}}$, the Eddington luminosity, we used the standard relation $L_{\rm{edd}} = 1.26\times10^{38}M_{\rm{BH}}$.

\begin{figure}[t!]
\begin{center}
\epsscale{1.0}
\plotone{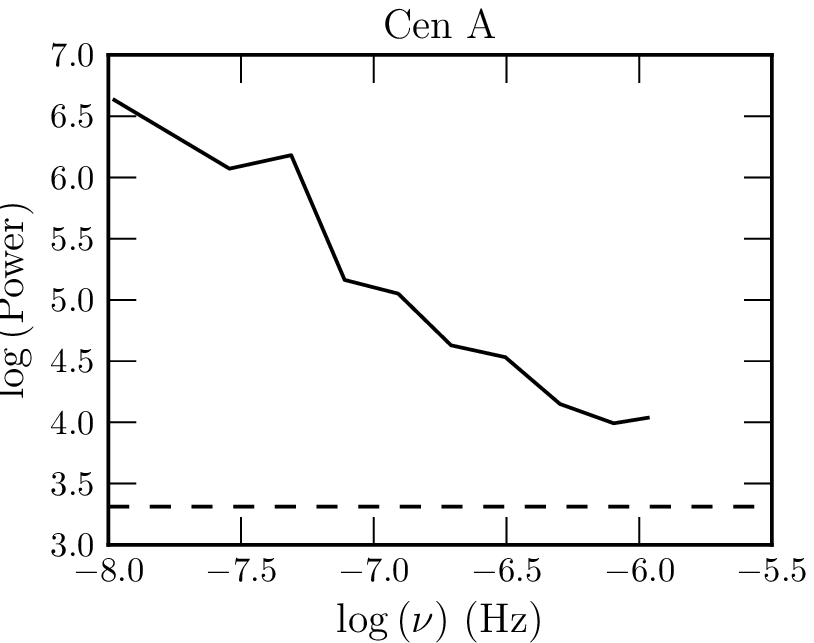}
\plotone{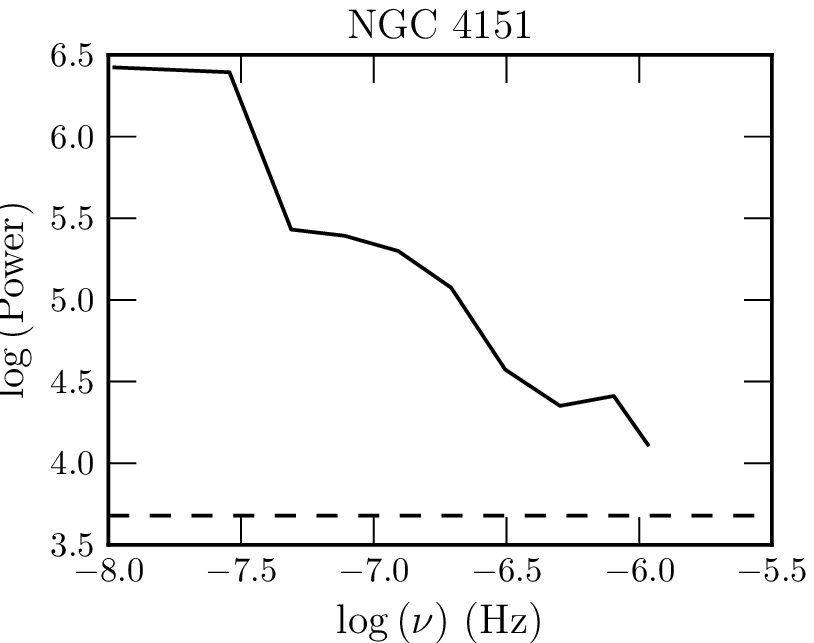}
\end{center}
\caption{Observed PSD's of Cen A and NGC 4151 constructed using Equation~\ref{eq_2} and the BAT lightcurves. Solid line is the PSD while the dashed line is the estimated white noise from Equation~\ref{eq_3}. The units on the power are $(\rm{rms/mean})^{2}(\rm{Hz})^{-1}$\label{fig_2}}
\end{figure}

Figure~\ref{fig_2} shows the observed PSDs for two of our brightest AGN, Cen A and NGC 4151 as well as their estimated Poisson level from equation~\ref{eq_3}. Most of the observed PSDs have a characteristic downward slope, but there is varying normalization. To quantify the best-fit parameters we use a technique based on Monte Carlo simulations that is described in the next section.

\subsection{Model Fitting}
 There are three main complications that arise when attempting to fit a model to an observed PSD. The first two are windowing effects known as {\it red-noise leak} and {\it aliasing}. 
 
 {\it Red-noise leak} is the transfer of power from frequencies lower than the minimum frequency sampled by the observed light curve (i.e. $T_{\rm obs}^{-1}$, $T_{\rm obs}$ is the length of the light curve) to higher frequencies. This occurs due to the finite length of the light curves, where we have essentially multiplied the intrinsic, infinitely long light curve by a box function with a width equal to the length of the observed light curve. 
 
 {\it Aliasing} is the transfer of power from frequencies higher than the maximum frequency sampled by the rebinned observed light curve (i.e. $(2\Delta T_{\rm rebin})^{-1}$ or the Nyquist frequency, $\Delta T_{\rm rebin}$ is the width of each time bin of the light curve) to lower frequencies and occurs due to the discrete sampling of the observed light curve. This can be thought of again as multiplying the continuously sampled intrinsic light curve by a series of box functions that have a width equal to the sampling time. Both of these effects multiply the true light curve by a window function in the time domain which is manifested as a convolution in the Fourier domain. Both are also model dependent effects, because the strength of each is dependent on the true variability power in the corresponding frequencies for each effect. {\it A priori} we have no prescription for correcting for them unless we assume a PSD shape. \citep[see][for a more detailed discussion.]{1993MNRAS.261..612P,2003MNRAS.345.1271V} Aliasing however probably isn't as big of an effect on the BAT PSD's since we are averaging over at least 20 points for each time bin. Aliasing only becomes a large effect if there are big differences between the exposure of the time bin and the gaps between the time bins (i.e. a 1 ksec observation every 500 ksec).

The third complication is the determination of the errors in the raw periodograms. \citet{1993MNRAS.261..612P} found that only when there are $>20$ points averaged together in a PSD logarithmic bin does the estimate of the power at a given frequency follow a Gaussian distribution.  High frequency bins easily meet this criterion however the low frequency points in the raw PSDs contain much less than $20$ points so errors cannot be determined in the usual way by calculating the scatter in a bin. 

To account for both windowing effects and the error determination, we use a Monte Carlo based technique first developed by \citet*{2002MNRAS.332..231U}, hereafter UMP02, called PSRESP that stands for Power Spectral Response, and is a method that improved upon the response method introduced by \citet{1992ApJ...400..138D}. The basic process is to simulate many light curves for each set of model parameters, calculate and average together the PSDs of the simulated light curves, and determine the goodness of fit for the model parameters which is defined by a rejection probability. The set of parameters with the lowest rejection probability are then deemed the best-fit model parameters. In the next sections we detail each step of the model fitting.

\subsubsection{Simulating Light Curves}\label{sim_lc}
To simulate the light curves we use the algorithm described in \citet{1995A&A...300..707T}. This algorithm not only randomizes the amplitude but also the phase which more realistically simulates the noise process that produced the observed PSD. The resulting simulated light curve is then calculated using the inverse Fourier Transform.

The first step is to input a model PSD that has some continuous shape that we wish to fit to the observed PSD. The normalization in the model is arbitrary since it will just multiply through and can be determined later in the process. We must also choose for the simulated light curve a time resolution, $\Delta T_{\rm sim}$ and overall length, $T_{\rm sim}$. To fully account for aliasing, we chose $\Delta T_{\rm sim}=800$s since this is the average exposure time for the raw BAT light curves. Any variability from timescales less than 800s are averaged over and won't contribute to aliasing. For the overall length of the simulated light curve we chose 1000$T_{\rm obs}$. In this way the simulated long light curve can be broken up into 1000 simulated short light curves that are $T_{\rm obs}$ long. Each short light curve will have taken into account red-noise leak since it was originally produced from a PSD with nonzero power at frequencies lower $(T_{\rm obs})^{-1}$.

Each short simulated curve is then resampled using the following procedure. Only those simulated light curve points that correspond to a time that was sampled in the unbinned observed light curve are used. All others that don't are thrown away. This produces a simulated light curve that best matches the observed light curve's sampling so all possible effects including gaps can be included modeled. The resampled simulated light curves are then rebinned and gaps are linearly interpolated through in the same way as the observed light curve. The PSD for each simulated light curve is calculated using equations~\ref{eq_1} \&~\ref{eq_2}.

\subsubsection{Determining Goodness of Fit}
After constructing the simulated PSD's, the logarithms of them are determined ($P_{\rm sim}$) and averaged together to form $P_{\rm avg}$ that represents our distorted model PSD and has all the effects of windowing and interpolation included. The errors in the power at each frequency, $\Delta P_{\rm sim}$, are equal to the rms spread in the set of $P_{\rm sim}$. $P_{\rm noise}$, which was calculated from Equation~\ref{eq_3} and is the linear noise power,  is then added in linear space as a constant to $P_{\rm avg}$.
\begin{equation}\label{eq_5}
P_{\rm avg+noise} = \log\left(10^{P_{\rm avg}} + P_{\rm noise}\right)
\end{equation}
 The test statistic, $\chi^2_{\rm dist}$, that is used to compare $P_{\rm obs}$ with $P_{\rm avg}$ from our model is defined in UMP02 as
\begin{equation}\label{eq_6}
\chi^2_{\rm dist} = \sum_{i}\frac{\left(P_{\rm avg+noise}(\nu_i) - P_{obs}(\nu_i)\right)^2}{\Delta P_{\rm sim}^2}
\end{equation}
where $\nu_i$ are the binned frequencies the PSDs are defined at.  The best-fitting normalization for the model is then determined by minimizing $\chi^2_{\rm dist}$. 

Although $\chi^2_{\rm dist}$ is defined in exactly the same way as the traditional $\chi^2$ statistic, it does not follow the same distribution. In fact $\chi^2_{\rm dist}$'s distribution changes depending on the input model parameters. Therefore to determine the goodness of fit we use the 1000 $P_{\rm sim}$  to calculate 1000 $\chi^2_{\rm dist,sim}$ which is defined the same as equation~\ref{eq_6} except that $P_{\rm obs}$ is replaced by $P_{\rm sim}$. Each $\chi^2_{\rm dist,sim}$ is minimized by finding the best-fit normalization in the same way as for the observed PSD. The goodness of fit is represented by the rejection probability, $R$, which is the percentile of $\chi^2_{\rm dist,sim}$ that are less than $\chi^2_{\rm dist}$. $R$ can be thought of as analogous to a p-value for a test statistic that measures the probability that the test statistic occurred by chance. A high $R$ (i.e. low p-value) signifies that the model does not correctly describe the observed PSD. For this paper however we will report Acceptance Probability which is just $1-R$. (See UMP02 for a full description of the fitting procedure.)

\section{Results}\label{sect_4}
\subsection{Unbroken Power Law}
For each source we first attempted to fit an unbroken power law model which has the form
\begin{equation}\label{eq_7}
P(\nu) = A_0\left(\frac{\nu}{\nu_0}\right)^{-\alpha}
\end{equation}
$A_0$ is the normalization and we chose $\nu_0$ to be equal to $10^{-6}$ Hz to be consistent with previous PSD studies of AGN. $\alpha$ is the power law slope which we varied from $\alpha=0-3.0$ with a step size of 0.05, because none of the observed PSDs show any evidence for a positive slope or a slope greater than 3. The best fit parameters for each of the AGN are listed in Table~\ref{tbl_3} and plots of the best fitting models with the observed PSDs can be seen in Figures~\ref{fig_3}. Errors on $\alpha$ were determined by calculating the amount $\alpha$ needs to change to increase $R$ by 1$\sigma$ assuming a Gaussian distribution.
\begin{equation}\label{eq_8}
\sigma = \sqrt{2}\mathrm{erf}^{-1}(R)
\end{equation}
\begin{equation}\label{eq_9}
R_{\rm crit} =  \mathrm{erf}\left(\frac{\sigma+1}{\sqrt{2}}\right)
\end{equation}
The confidence interval can then be determined by where $R(\alpha)=R_{\rm crit}$. This method was first used by \citet{2003ApJ...593...96M}, and it has been shown that the errors determined in this way correspond to a 99\% confidence interval when an unbroken power law model is used and a 90\% confidence interval when a singly broken power law model is used \citep[Appendix B]{2010ApJ...724...26M}. The errors on $A_0$ were calculated from the rms spread of the 1000 best fitting normalizations found when producing the $\chi^2_{\rm dist}$ distribution. Also instead of showing rejection probability, we show the acceptance probability which is just $1-R$.

\begin{figure*}[htbp]
\centering
\includegraphics[width = 0.8\textwidth]{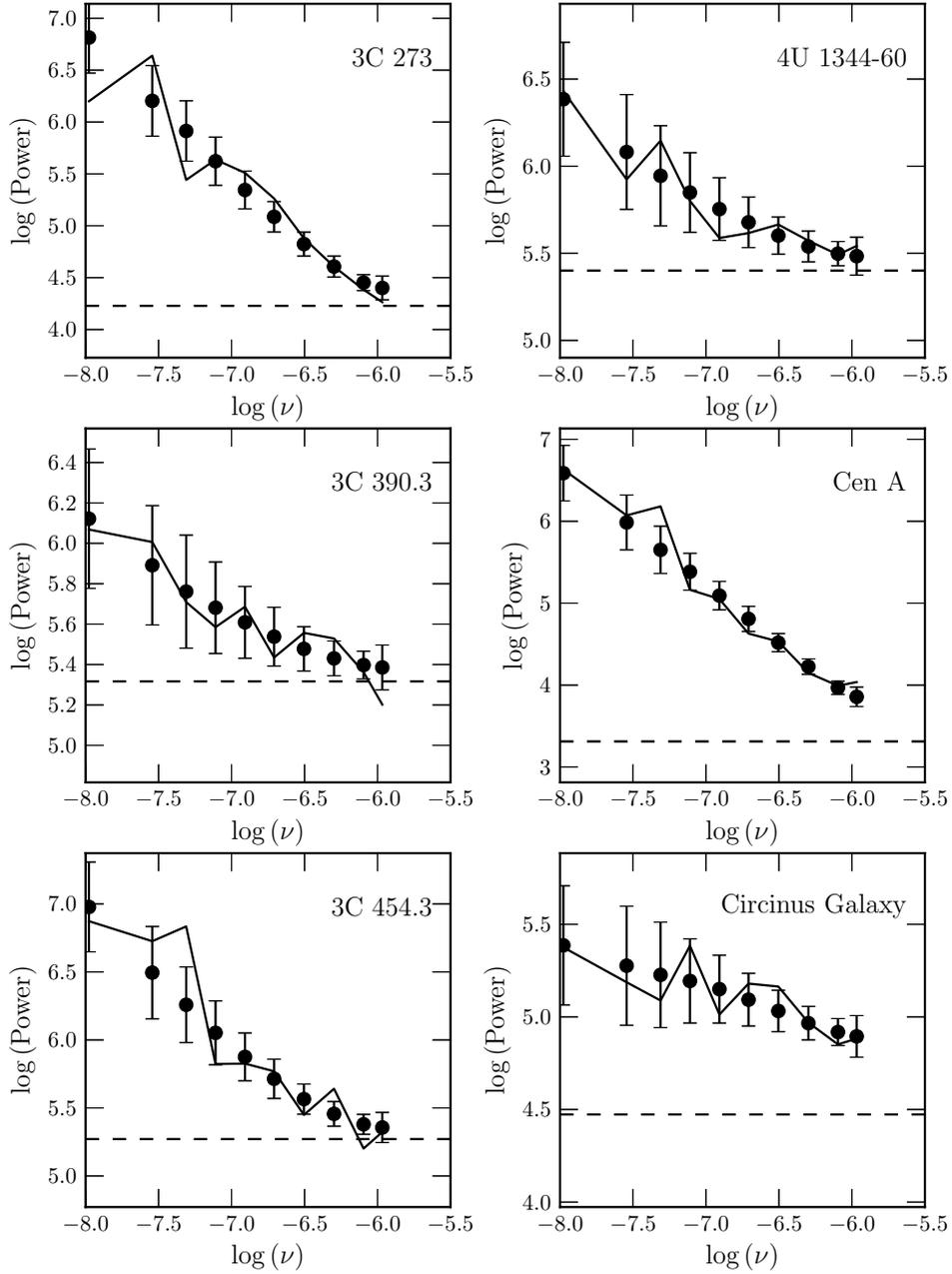}
\caption{Observed PSD's (solid line) plotted with best fitting distorted model PSD (dots with error bars). The distorted model contains all windowing effects and error bars are 1$\sigma$ uncertainties. Dashed line indicates estimated white noise level for each AGN due to measurement error. See Table~\ref{tbl_3} for a list of best fitting parameters and acceptance probabilities.\label{fig_3}}
\end{figure*}

\begin{figure*}[htbp]
\centering
\figurenum{~\ref{fig_3}}
\centering
\includegraphics[width = 0.8\textwidth]{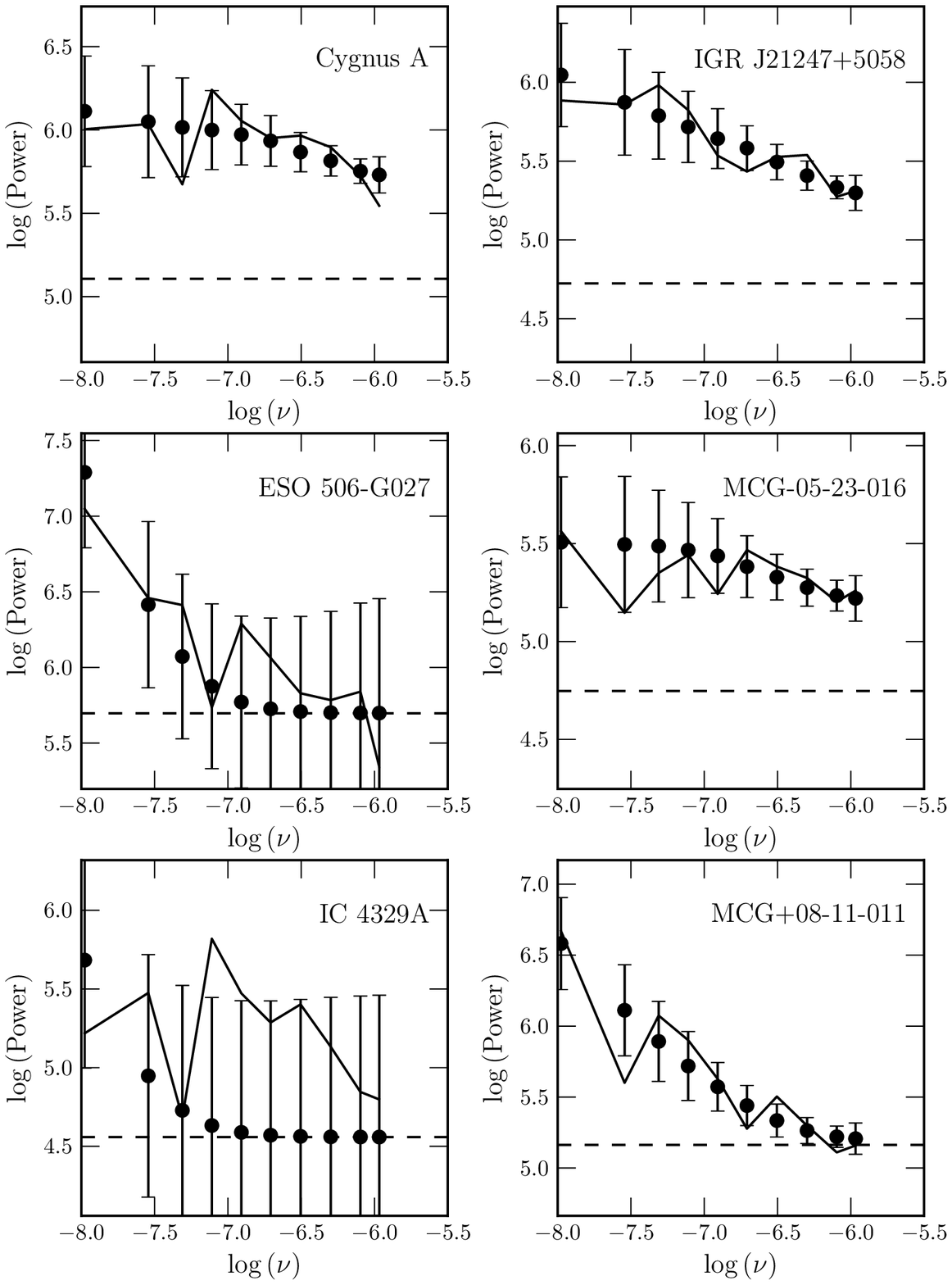}
\caption{continued}
\end{figure*}

\begin{figure*}[htbp]
\figurenum{~\ref{fig_3}}
\centering
\includegraphics[width = 0.8\textwidth]{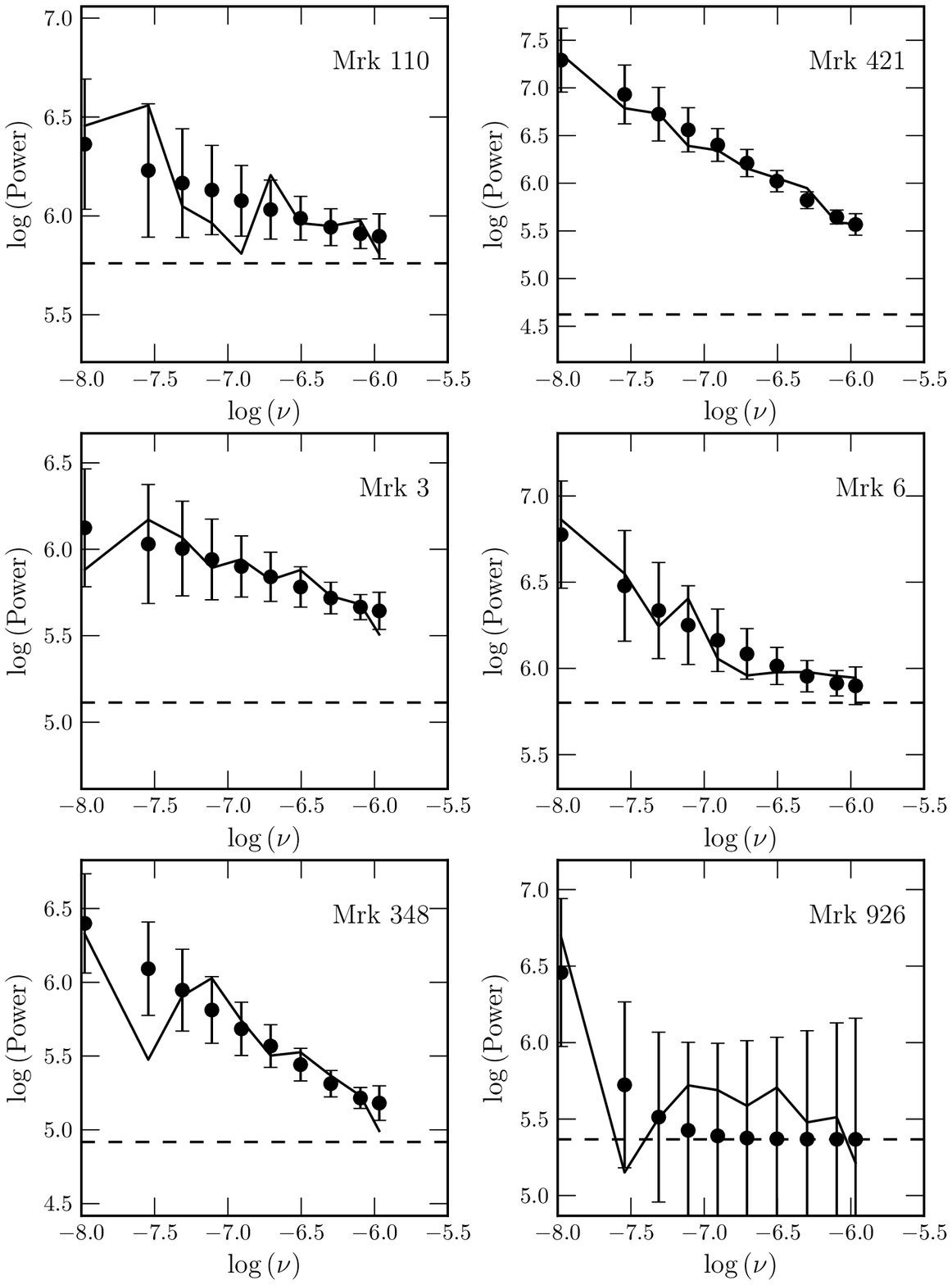}
\caption{continued}
\end{figure*}

\begin{figure*}[htbp]
\figurenum{~\ref{fig_3}}
\centering
\includegraphics[width = 0.8\textwidth]{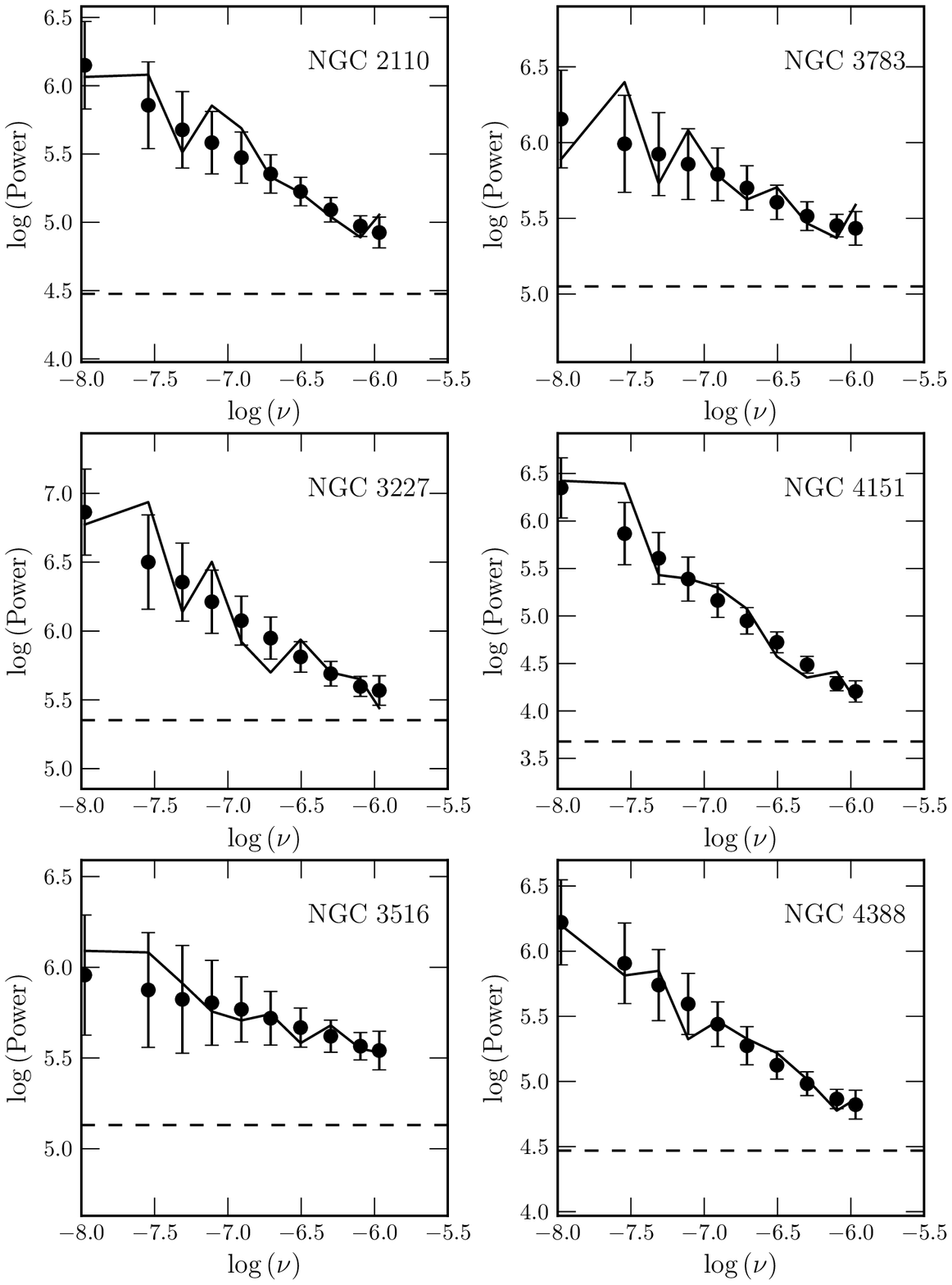}
\caption{continued}
\end{figure*}

\begin{figure*}[htbp]
\figurenum{~\ref{fig_3}}
\centering
\includegraphics[width = 0.8\textwidth]{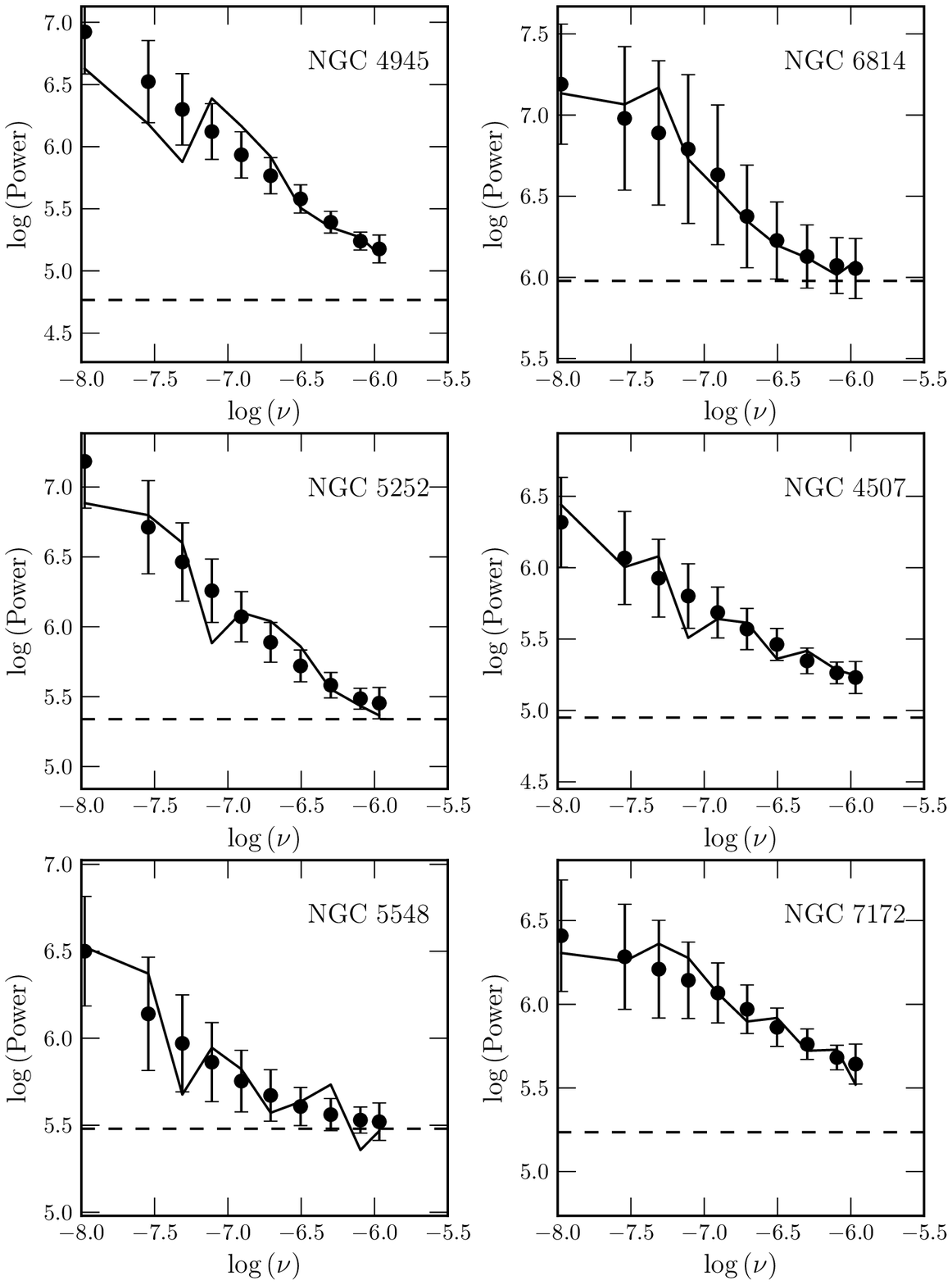}
\caption{continued}
\end{figure*}

\begin{deluxetable}{llcc}[t!]
\tablecaption{PSD Best Fit Parameters for Unbroken Model\label{tbl_3}}
\tablewidth{0pt}
\tablehead{\colhead{} & \colhead{} & \colhead{$\log A_0 $} & \colhead{} \\
\colhead{Name} & \colhead{$\alpha$} & \colhead{$(\rm{Hz}^{-1})$} & \colhead{$1-R$}\\
\colhead{(1)} & \colhead{(2)} & \colhead{(3)} & \colhead{(4)}}
\startdata
3C 273 & 1.35$^{+0.35}_{-0.4}$   & 4.06$\pm$0.05  & 0.170 \\
3C 390.3 & 0.75$^{+0.65}_{-0.55}$  & 4.53$\pm$0.04  & 0.728 \\
3C 454.3 & 1.10$^{+\infty}_{-0.3}$   & 4.73$\pm$0.04  & 0.064 \\
4U 1344-60 & 0.80$^{+0.6}_{-0.65}$  & 4.71$\pm$0.04  & 0.979 \\
Cen A & 1.35$^{+0.35}_{+0.35}$  & 3.82$\pm$0.06  & 0.379 \\
Circinus Galaxy & 0.40$^{+0.35}_{-\infty}$   & 4.38$\pm$0.04  & 0.897 \\
Cygnus A & 0.30$^{+0.35}_{-\infty}$   & 5.20$\pm$0.05  & 0.563 \\
ESO 506-G027 & 2.40$^{+\infty}_{-0.8}$   & 2.06$\pm$0.91  & 0.164 \\
IC 4329A & 0.50$^{+0.35}_{-0.35}$  & 4.52$\pm$0.05  & 0.009 \\
IGR J21247+5058 & 0.50$^{+0.25}_{-\infty}$   & 4.97$\pm$0.04  & 0.810 \\
MCG-05-23-016 & 0.15$^{+0.4}_{-\infty}$  & 4.34$\pm$0.05  & 0.936 \\
MCG+08-11-011 & 1.10$^{+0.6}_{-0.45}$  & 4.32$\pm$0.04  & 0.365 \\
Mrk 110 & 0.50$^{+0.45}_{-\infty}$   & 5.15$\pm$0.04  & 0.626 \\
Mrk 3 & 0.40$^{+0.3}_{-\infty}$   & 5.16$\pm$0.04  & 0.956 \\
Mrk 348 & 0.75$^{+0.3}_{-0.35}$  & 4.85$\pm$0.04  & 0.473 \\
Mrk 421 & 0.85$^{+0.25}_{-0.25}$  & 5.56$\pm$0.04  & 0.920 \\
Mrk 6 & 0.70$^{+0.7}_{-0.5}$   & 5.20$\pm$0.04  & 0.986 \\
Mrk 926 & 2.45$^{+\infty}_{-0.85}$  & 1.06$\pm$1.03  & 0.146 \\
NGC 2110 & 0.70$^{+0.35}_{-0.2}$  & 4.69$\pm$0.04  & 0.674 \\
NGC 3227 & 0.80$^{+0.4}_{-0.35}$  & 5.20$\pm$0.05  & 0.292 \\
NGC 3516 & 0.35$^{+0.4}_{-\infty}$  & 5.00$\pm$0.05  & 0.995 \\
NGC 3783 & 0.50$^{+0.3}_{-\infty}$   & 4.99$\pm$0.05  & 0.497 \\
NGC 4151 & 1.10$^{+0.45} _{-0.25}$ & 4.13$\pm$0.04  & 0.205 \\
NGC 4388 & 0.80$^{+0.35}_{-0.3}$   & 4.59$\pm$0.04  & 0.874 \\
NGC 4945 & 0.95$^{+0.25}_{-0.35}$  & 5.00$\pm$0.04  & 0.406 \\
NGC 5252 & 1.10$^{+0.35}_{-0.4}$   & 4.95$\pm$0.04  & 0.544 \\
NGC 5548 & 0.95$^{+\infty}_{-0.45}$  & 4.53$\pm$0.04  & 0.241 \\
NGC 6814 & 0.85$^{+0.65}_{-\infty}$  & 5.23$\pm$0.18  & 0.993 \\
NGC 4507 & 0.70$^{+0.3}_{-0.4}$   & 4.86$\pm$0.04  & 0.912 \\
NGC 7172 & 0.50$^{+0.35}_ {-\infty}$   & 5.27$\pm$0.05  & 0.967
\enddata
\tablecomments{Col. (1): Name of AGN. Col. (2): Best fitting power law slope with 1$\sigma$ errors determined using the method described in Section~\ref{sect_4}. Col. (3): Best fitting normalization. Col. (4): Acceptance probability for the best fit. High values represent a good fit.}
\end{deluxetable}

\begin{figure}[t!]
\begin{center}
\epsscale{1.0}
\plotone{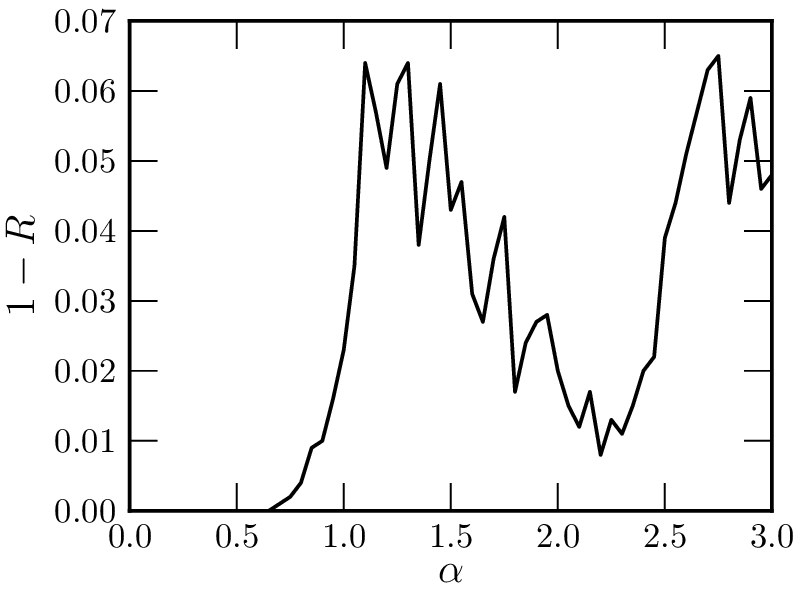}
\plotone{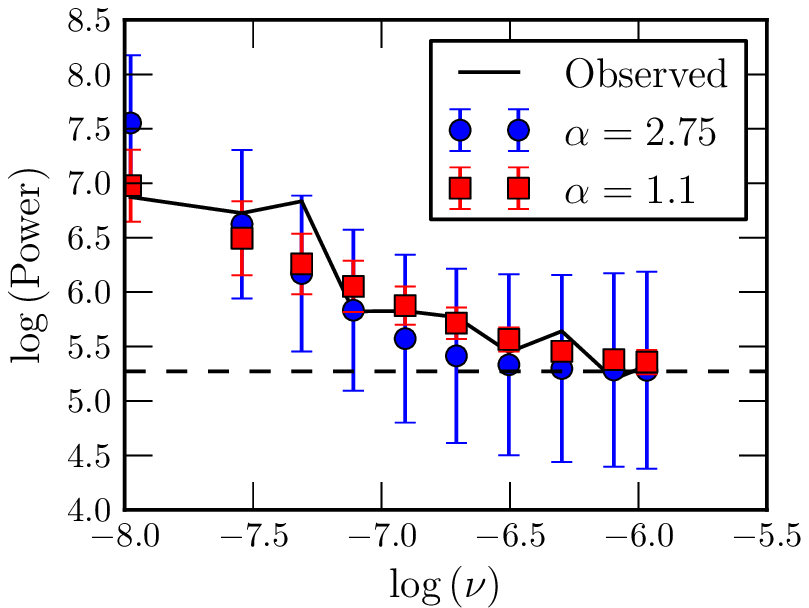}
\caption{Fit results for 3C 454.3. Top: Curve of Acceptance Probability against power law slope ($\alpha$). Notice the two peaks in the curve at 1.1 and 2.75. Bottom: Comparison of the fit with $\alpha = 1.1$ (squares, colored red in online version) and $\alpha = 2.75$ (circles, colored blue in online version). The black solid curve is the observed PSD and the dashed line is the estimated $P_{noise}$ level. \label{fig_8}}
\end{center}
\end{figure}

For 3C 454.3 two peaks in the Acceptance Probability occured, one at $\alpha = 1.1$ and the other at 2.75. Both had very nearly the same Acceptance Probability (.064 vs. .065 respectively) so we chose to use the $\alpha = 1.1$ fit based on visual inspection of the fits. It is well known that large values of $\alpha$ can produce large values of Acceptance Probability because red-noise leak will cause bigger error bars, and thus smaller values of $\chi^2_{\rm dist}$ \citep{2003MNRAS.339.1237V, 2003MNRAS.341..496V}. The $\alpha = 2.75$ fit is mainly dominated by $P_{\rm noise}$ and the values of the distorted model don't correspond well with the observed PSD values. Figure~\ref{fig_8} shows the two peaks in Acceptance Probability and the fits using both peaks. 

\begin{figure}[t!]
\epsscale{1.0}
\plotone{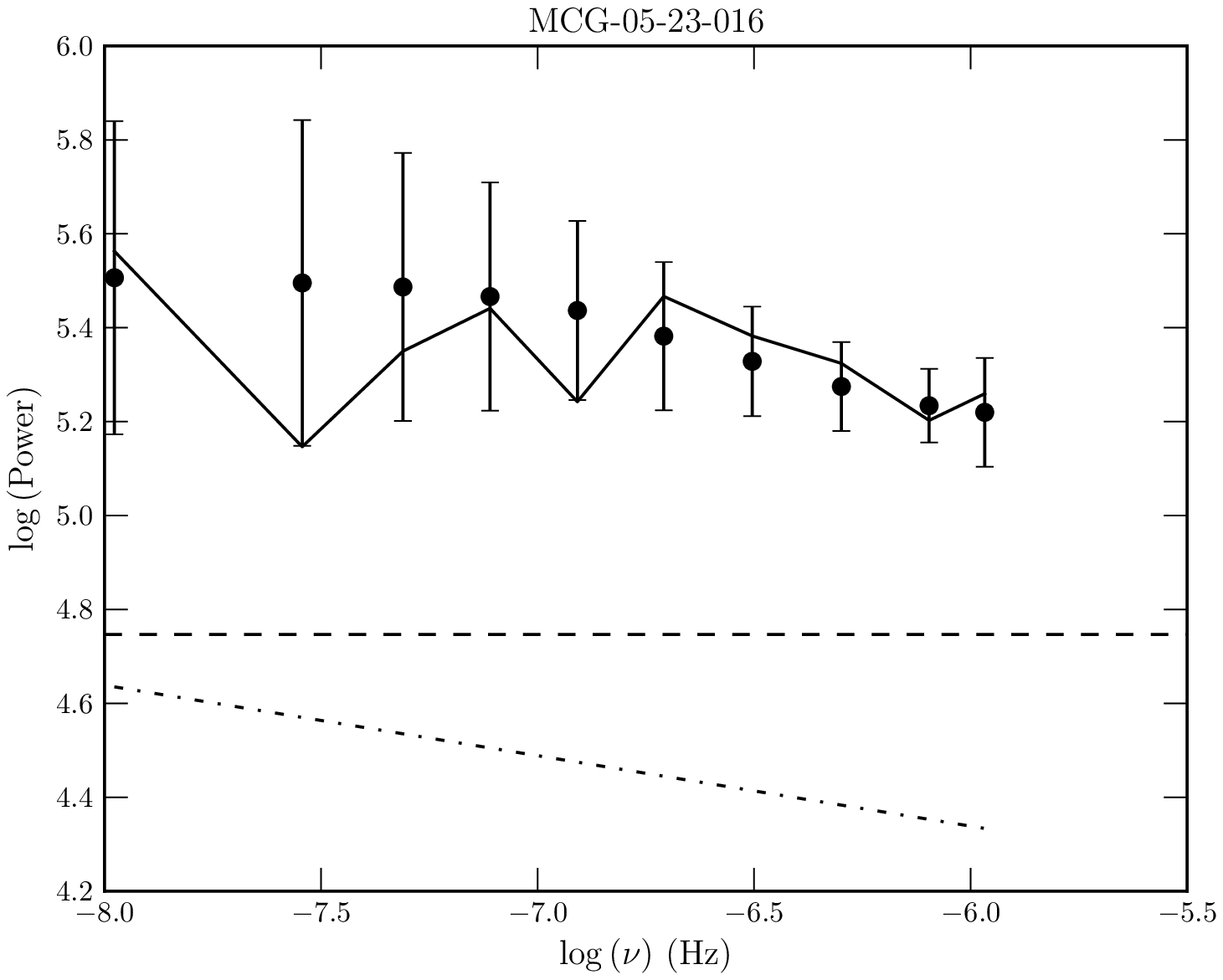}
\caption{Fit results for MCG-05-23-016 showing the model fit without distortion (dash-dot) together with $P_{noise}$ (dashed line), the distorted model fit (dots with errorbars) and observed PSD. The distorted model is completely dominated by $P_{noise}$.\label{fig_9}}
\end{figure}

We have identified 4 sources which represented ``bad'' fits because most of the power in the fitted model is due to $P_{\rm noise}$ shown by a severe flattening in the PSD. These are ESO 506-G027, MCG--05-23-016, IC 4329A, and Mrk 926. For ESO 506-G027, IC 4329A, and Mrk 926 it is clear from the figures the fit is bad but for MCG--05-23-016 a closer look was needed. Figure~\ref{fig_9} shows the observed PSD, model fit, and model fit removing any distortions, including window effects and $P_{\rm noise}$. From this, we can see that the fit is entirely dominated by $P_{\rm noise}$ because the undistorted model is below $P_{\rm noise}$ so we remove this source from any analysis. The reason for these bad fits is either a lack of intrinsic variability which is suspected for IC 4329A and MCG--05-23-016, or a high $P_{\rm noise}$  which is suspected for ESO 506-G027 and Mrk 926. $P_{\rm noise}$ may not be the same in the BAT data due to the method via which the catalog is constructed \citep{2010ApJS..186..378T} which involves cleaning the data. 

24 out of 26 sources (our original 30 sources minus the 4 with bad fits)  had Acceptance Probabilities greater than 20\% indicating the unbroken model is a sufficient model to characterize the hard X-ray PSDs of AGN. The two sources with less than 20\% acceptance are 3C 273 and 3C 454.3 which will later be tested with a singly broken model. 

The average power law slope  for all AGN with a best fit slope was 0.78$\pm$0.29 where the error is the standard deviation of the slopes. If we restrict the average to only those slopes that were fully constrained then it changes to 0.92$\pm$0.22. However this might be biased against flatter slopes because those are more likely to be unconstrained. 

16 out of 26 sources had their PSD power law slopes fully constrained. 10 of the other 14 sources only resulted in upper limits on the slope, and the other 4 unconstrained sources only resulted in lower limits on the slope. These are indicated in Table~\ref{tbl_3} by a $\pm\infty$ in one of the error bars. Flat slopes are difficult to constrain on the lower end because the effect of $P_{\rm noise}$ is to flatten the PSD out. So if the observed PSD power is near the level of $P_{\rm noise}$, many different slopes will provide a good fit. 

\subsection{Singly Broken Power Law}

\begin{figure}[t!]
\epsscale{1.0}
\plotone{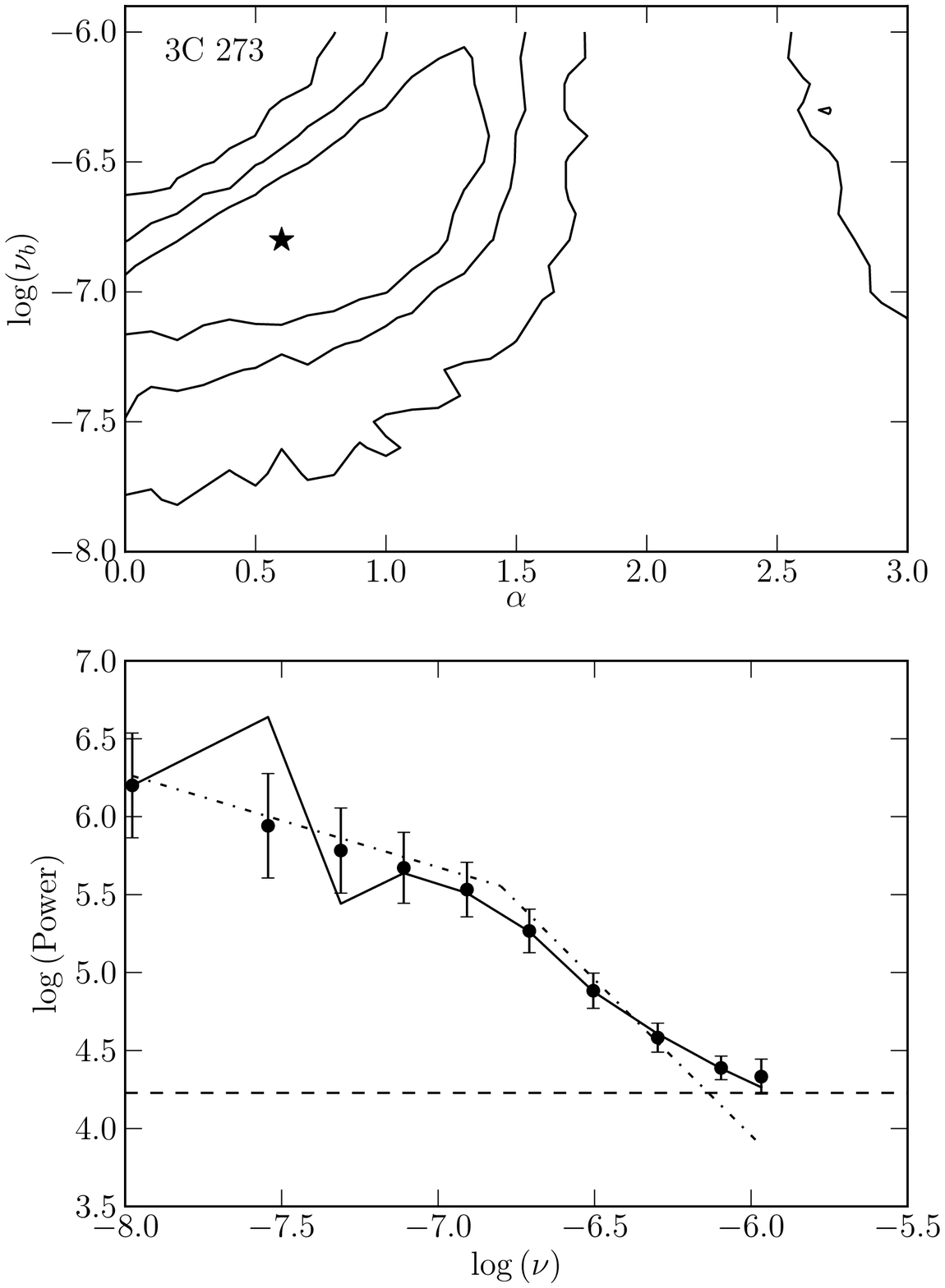}
\caption{Fit results for 3C 273 using the singly broken model described by Equation~\ref{single_broken}. Top: Contours of constant Acceptance Probability for the different pairs of break frequency ($\nu_b$) and low frequency slope ($\alpha$). 3 contours were drawn at 32, 10, and 1\% the peak of Acceptance Probability. Stars indicate the location of the best fit parameters. Bottom: Observed PSD (solid line) plotted along with best fit distorted model (dots with error bars) and undistorted model (dot-dash). Horizontal dashed line represents the Poisson noise level.\label{fig_15}}
\end{figure}

\begin{figure}[t!]
\epsscale{1.0}
\plotone{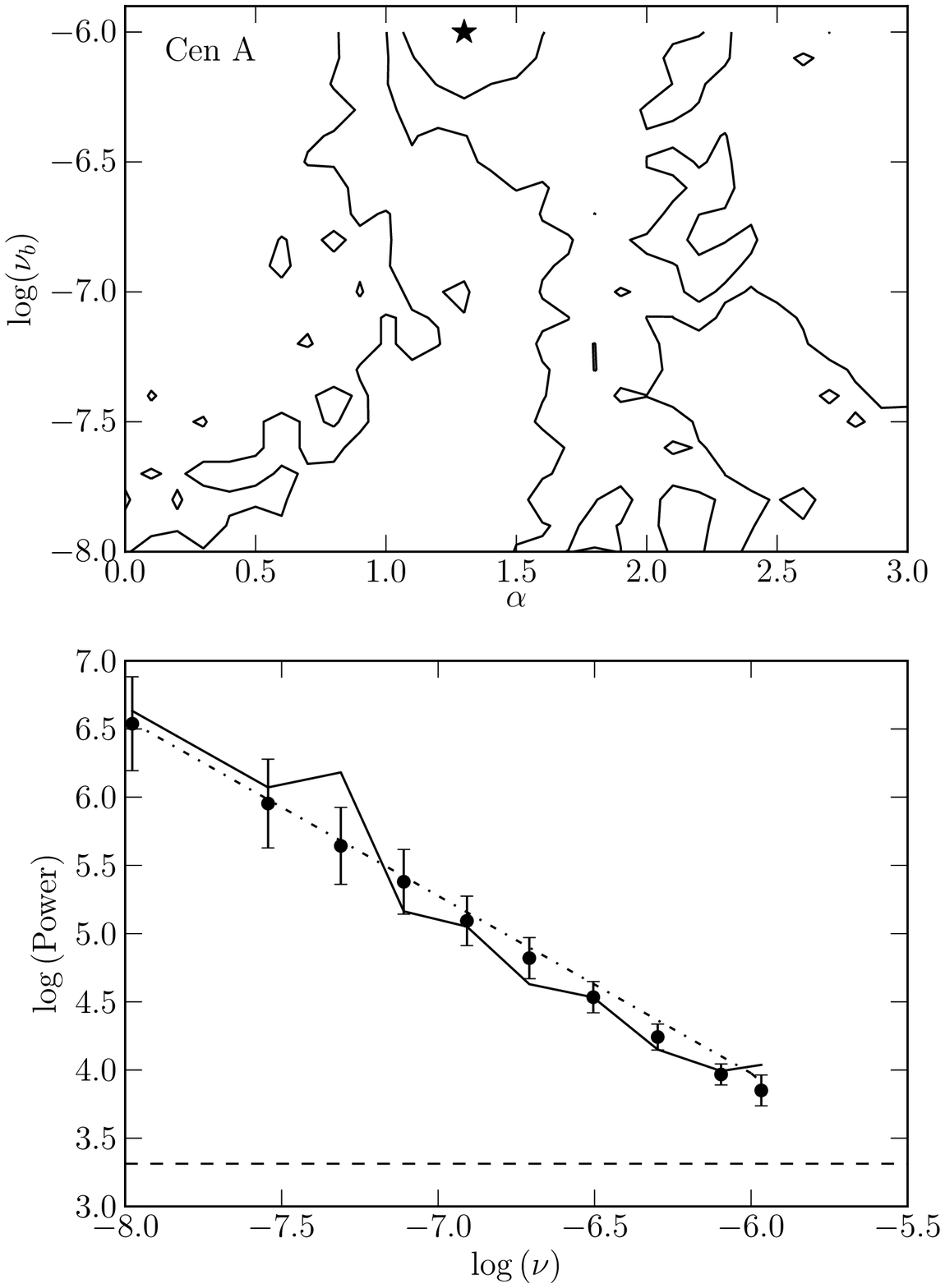}
\caption{Same as Figure~\ref{fig_15} but for Cen A\label{fig_16}}
\end{figure}

\begin{figure}[t!]
\epsscale{1.0}
\plotone{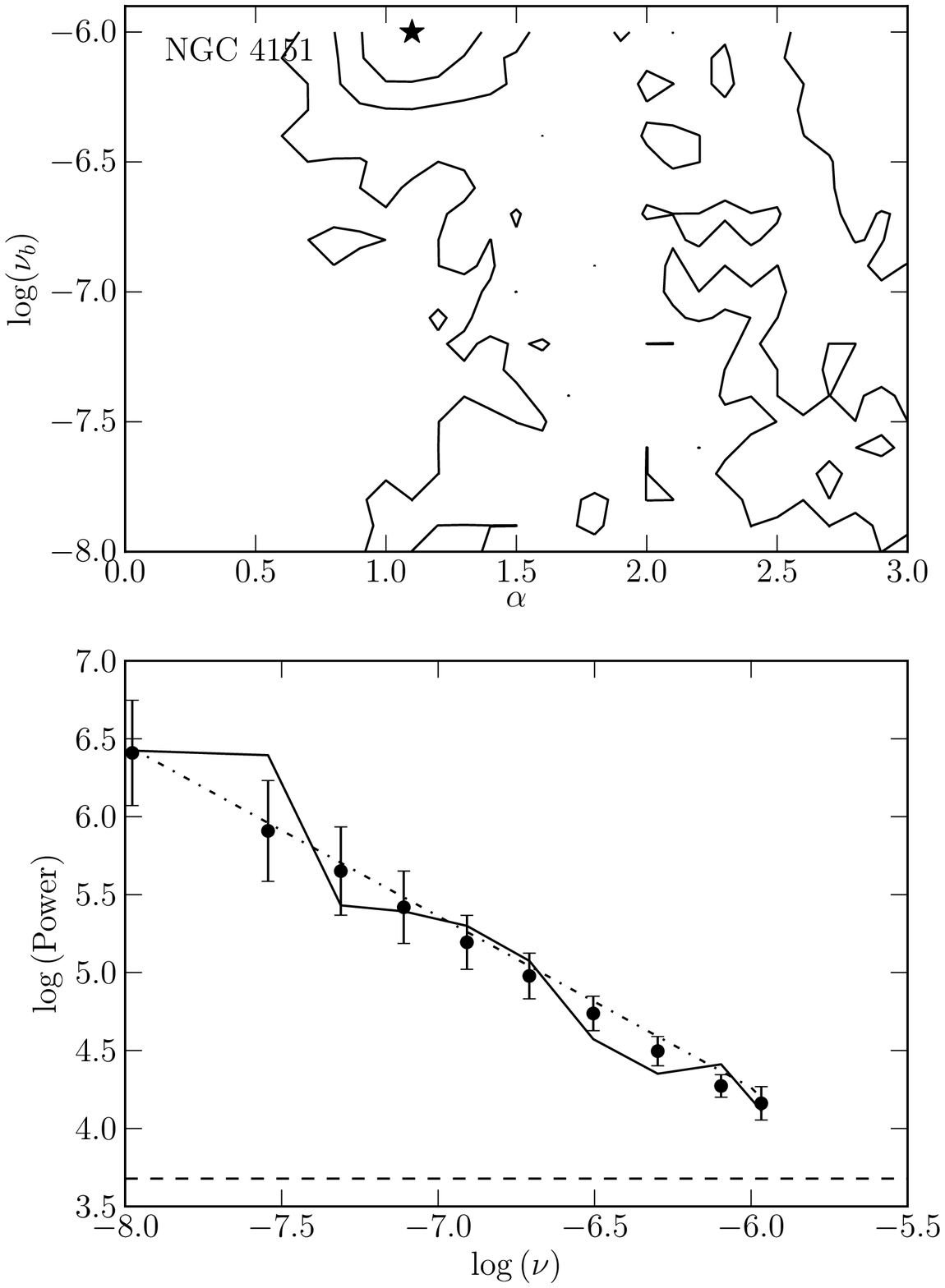}
\caption{Same as Figure~\ref{fig_15} but for NGC 4151\label{fig_17}}
\end{figure}

\begin{figure*}[t!]
\centering
\includegraphics[width = 0.8\columnwidth]{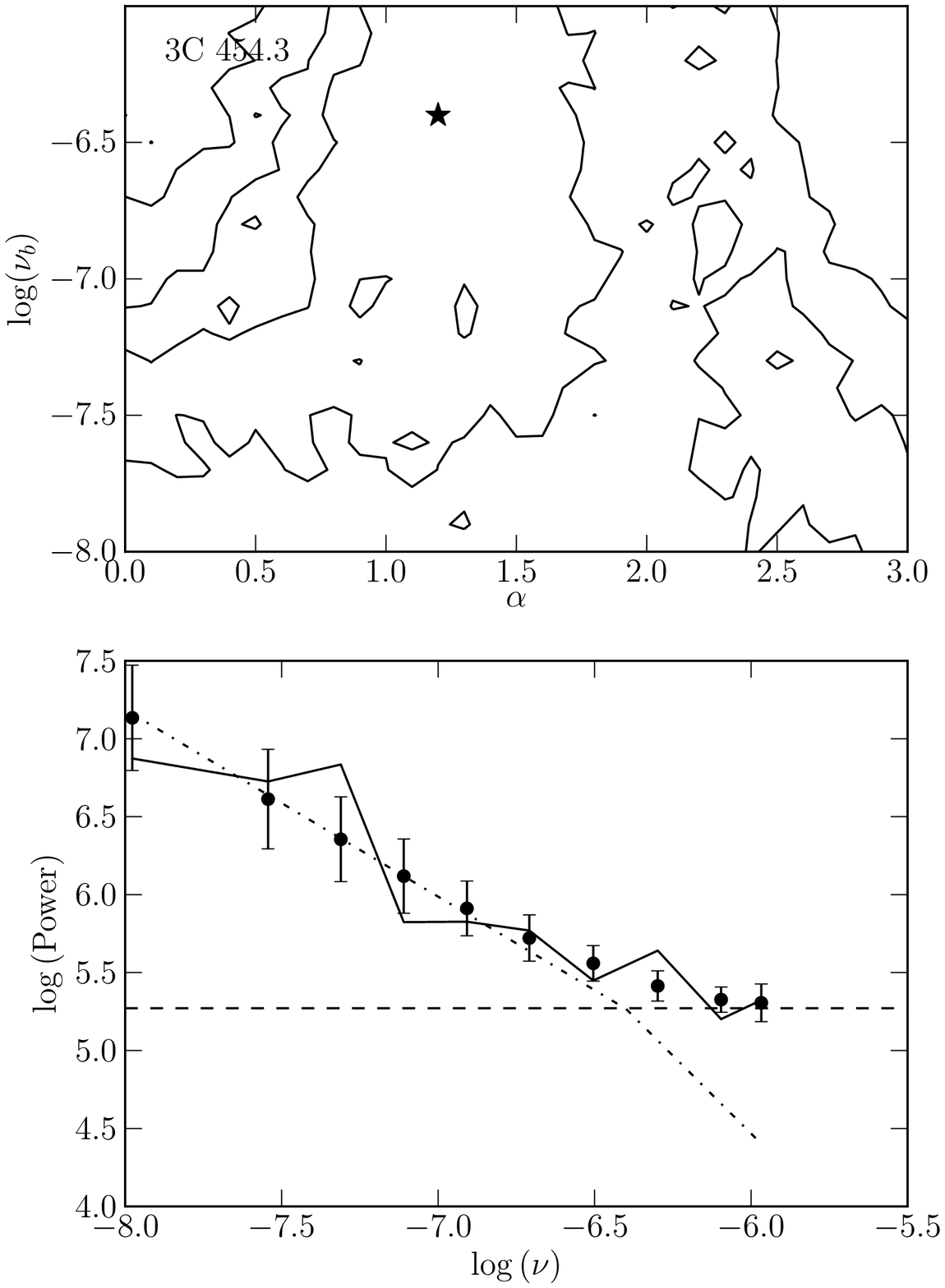}
\includegraphics[width = 0.8\columnwidth]{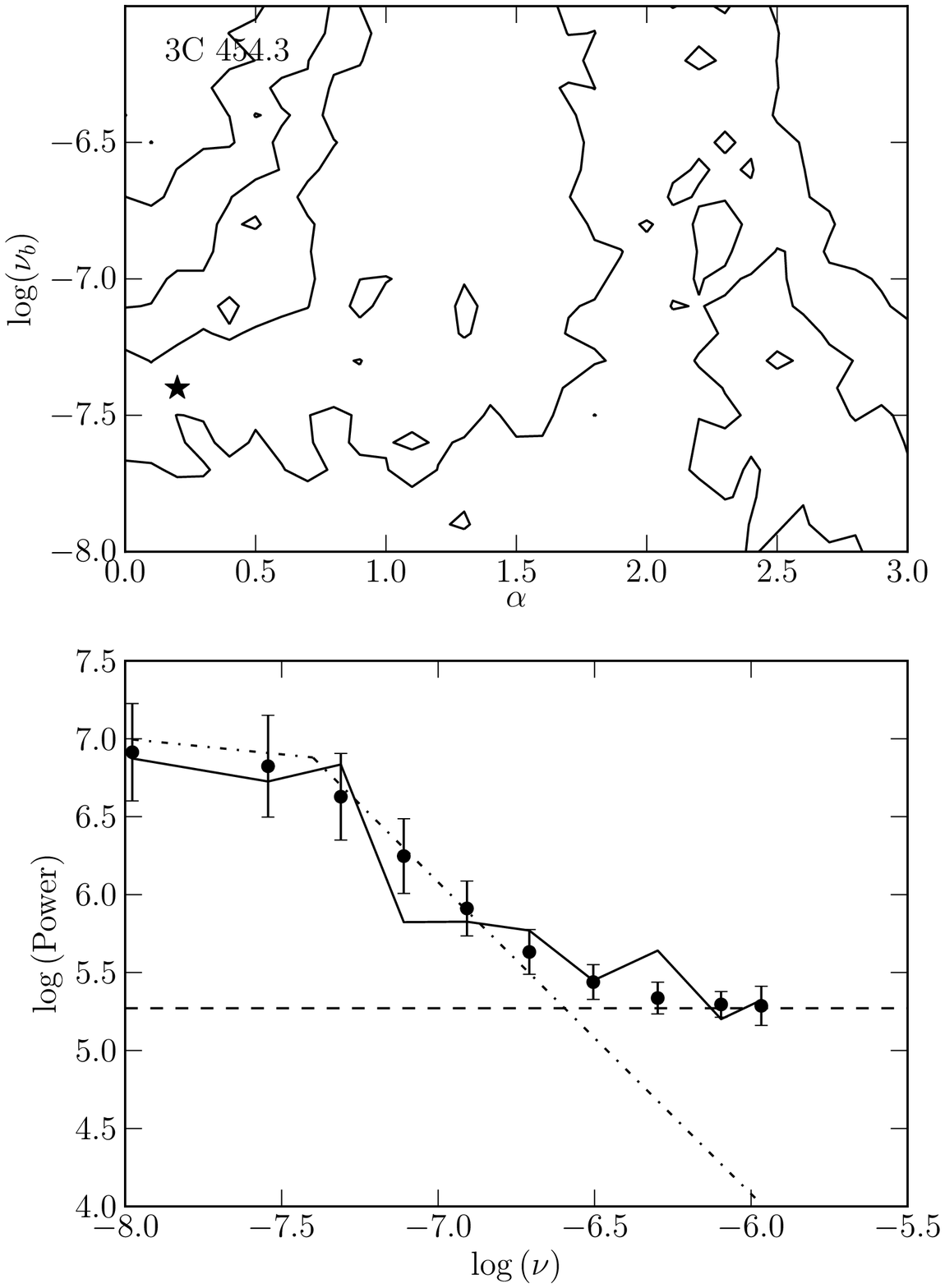}
\caption{Same as Figure~\ref{fig_15} but for 3C 454.3. The left plot shows the best fit PSD with a break frequency of $\log(\nu_b) = 6.4$ and power law slope $\alpha = 1.2$. The right plot shows the second best fit PSD with a break frequency of $\log(\nu_b) = 7.4$ and power law slope $\alpha = 0.2$. \label{fig_C}}
\end{figure*}

\begin{deluxetable*}{llcccc}[t!]
\tabletypesize{\scriptsize}
\tablewidth{0pt}
\tablecaption{PSD Best Fit Parameters for Singly Broken Model\label{tbl_7}}
\tablehead{\colhead{} & \colhead{} & \colhead{} & \colhead{$\log\nu_b$} & \colhead{$\log A_{1}$} &\colhead{} \\
\colhead{Name} & \colhead{Type} & \colhead{$\alpha$} & \colhead{(Hz)} & \colhead{(Hz$^{-1}$)} & \colhead{$1-R$} \\
\colhead{(1)} & \colhead{(2)} & \colhead{(3)} & \colhead{(4)} & \colhead{(5)} & \colhead{(6)}}
\startdata
Cen A        & NLRG & 1.3$^{+0.6}_{-0.3}$ & -6.0$^{+\infty}_{-0.4}$  & 3.97$\pm$0.05 & 0.262 \\
3C 273	& Blazar & 0.6$^{+0.7}_{-\infty}$ & -6.8$^{+0.3}_{-0.4}$  & 5.55$\pm$0.02 & 0.680 \\
NGC 4151 & Sy 1.5  & 1.1$^{+0.5}_{-0.3}$  & -6.0$^{+\infty}_{-0.3}$ & 4.26$\pm$0.04 & 0.203 \\
3C 454.3 & Blazar & 1.2$^{+\infty}_{-0.8}$ & -6.4$^{+\infty}_{-\infty}$ & 5.27$\pm$0.05 & 0.096 \\
3C 390.3 & Blazar & 1.0 & -6.2$^{+\infty}_{-1.2}$ & 4.64$\pm$0.04 & 0.621 \\
4U 1344-60 & Sy 1.5 & 1.0 & -6.0$^{+\infty}_{-1.4}$ & 4.64$\pm$0.04 & 0.802 \\
Circinus Galaxy & Sy 2 & 1.0 & -6.2$^{+\infty}_{-1.2}$ & 4.64$\pm$0.04 & 0.621 \\
Cygnus A & NLRG & 1.0 & -8.0$^{+\infty}_{-\infty}$ & 7.70$\pm$0.13 & 0.000 \\
ESO 506-G027 & Sy 2 & 1.0 & -8.0$^{+\infty}_{-\infty}$ & 7.46$\pm$0.13 & 0.030 \\
IC 4329A & Sy 1 & 1.0 & -6.2$^{+0.1}_{-0.1}$ & 4.84$\pm$0.04 & 0.001 \\
IGR J21247+5058 & BLRG & 1.0 & -8.0$^{+\infty}_{-\infty}$ & 7.58$\pm$0.13 & 0.000 \\
MCG-05-23-016 & Sy 1.9 & 1.0 & -8.0$^{+\infty}_{-\infty}$ & 6.40$\pm$0.13 & 0.000 \\
MCG+08-11-011 & Sy 1.9 & 1.0 & -6.5$^{\infty}_{-0.9}$ & 5.10$\pm$0.05 & 0.453 \\
Mrk 110 & Sy 1 & 1.0 & -6.0$^{+\infty}_{-\infty}$ & 4.91$\pm$0.04 & 0.073 \\
Mrk 3 & Sy 2 & 1.0 & -8.0$^{+\infty}_{-\infty}$ & 7.57$\pm$0.13 & 0.000 \\
Mrk 348 & Sy 2 & 1.0 & -6.0$^{+\infty}_{-0.5}$ & 4.88$\pm$0.04 & 0.073 \\
Mrk 421 & Blazar & 1.0 & -6.0$^{+\infty}_{-0.3}$ & 5.66$\pm$0.05 & 0.337 \\
Mrk 6 & Sy 1.5 & 1.0 & -6.0$^{+\infty}_{-1.9}$ & 5.06$\pm$0.04 & 0.786 \\
Mrk 926 & Sy 1.5 & 1.0 & -6.4$^{+\infty}_{-\infty}$ & 4.83$\pm$0.04 & 0.027 \\
NGC 2110 & Sy 2 & 1.0 & -6.0$^{+\infty}_{-0.4}$ & 4.76$\pm$0.04 & 0.085 \\
NGC 3227 & Sy 1.5 & 1.0 & -6.0$^{+\infty}_{-0.5}$ & 5.22$\pm$0.04 & 0.124 \\
NGC 3516 & Sy 1.5 & 1.0 & -6.0$^{+\infty}_{-0.1}$ & 5.05$\pm$0.04 & 0.001 \\
NGC 3783 & Sy 1.5 & 1.0 & -6.0$^{+\infty}_{-0.1}$ & 5.04$\pm$0.04 & 0.003 \\
NGC 4388 & Sy 2 & 1.0 & -6.0$^{+\infty}_{-0.4}$ & 4.65$\pm$0.04 & 0.444 \\
NGC 4507 & Sy 2 & 1.0 & -6.0$^{+\infty}_{-0.5}$ & 4.88$\pm$0.04 & 0.080 \\
NGC 4945 & Sy 2 & 1.0 & -6.0$^{+\infty}_{-0.4}$ & 5.13$\pm$0.04 & 0.263 \\
NGC 5252 & Sy 2 & 1.0 & -6.4$^{+\infty}_{-0.5}$ & 5.60$\pm$0.05 & 0.769 \\
NGC 5548 & Sy 1.5 & 1.0 & -6.5$^{+\infty}_{-\infty}$ & 5.12$\pm$0.05 & 0.258 \\
NGC 6814 & Sy 2 & 1.0 & -6.1$^{+\infty}_{-1.5}$ & 5.38$\pm$0.18 & 0.954 \\
NGC 7172 & Sy 2 & 1.0 & -6.1$^{+\infty}_{-0.1}$ & 5.46$\pm$0.04 & 0.005 \\
\enddata
\tablecomments{Results from fitting a singly broken model. For Cen A, NGC 4151,  3C 273, and 3C 454.3. $\alpha$, the power law slope below the break frequency and $\nu_b$, the break frequency were freely allowed to vary while the high frequency slope was fixed at 2. For the rest of the sources $\alpha$ was fixed at 1.0 and only $\nu_b$ was allowed to vary.} 
\end{deluxetable*}

For 4 AGN (3C 273, Cen A,  NGC 4151, and 3C 454.3)  we tested a singly broken power law model with the form
\begin{equation}\label{single_broken}
P(\nu) = \left\{
\begin{array}{rl}
A_1\left(\frac{\nu}{\nu_b}\right)^{-\alpha} & \nu \leq \nu_b \\
A_1\left(\frac{\nu}{\nu_b}\right)^{-\beta} & \nu \geq \nu_b.
\end{array} \right.
\end{equation}
where $A_1$ is the normalization, $\nu_b$ the frequency at which the PSD changes from a high frequency power law slope of $-\beta$ to a low frequency slope of $-\alpha$. We chose these AGN due to their steep best fit slope from the unbroken model and/or their low Acceptance Probability. We suspected that possibly our low frequency resolution for the PSD could be smoothing out a potential break frequency thus giving an ``average'' best fit slope somewhere between 1 and 2 depending on where the break occurs. 

Figures~\ref{fig_15}, ~\ref{fig_16}, ~\ref{fig_17}, ~\ref{fig_C}, and Table~\ref{tbl_7} show the results from this model. Fitting the unbroken power law model took on average 1 day per source on a machine with 3.0 GHz processing speed and 8 cores. Using a singly broken power law model with 2 more free parameters could take on the order of months, so to reduce the computational time, we froze $\beta=2$ reducing the number of free parameters to 2, $\nu_b$ and $\alpha$. Freezing the high frequency power law slope seems to be a valid decision since this is the typical value measured for the PSD in the 2--10 keV band \citep{2012A&A...544A..80G} ,and we are most likely to be sensitive to the low frequency slope rather than the high frequency one since we are probing the longest timescales. $\nu_b$ was stepped through every 0.1 in the log between -8 and -6, approximately the lowest and highest temporal frequencies sampled in our observed PSDs, and $\alpha$ was stepped through every 0.1 (as opposed to 0.05 with the unbroken model) between 0 and 3.

Both Cen A and NGC 4151 resulted in essentially the same fit as for the unbroken model. The best fit $\alpha$'s  were exactly the same as for the unbroken model (since we stepped through $\alpha$  with stepsize 0.1, 1.35 was not an option for Cen A) as well as the normalization. The best fit break frequencies also occured at the edge of the parameter space demonstrating that the break frequency must occur at a frequency outside the range sampled by the PSD's. 

The singly broken model only gave marginal improvement for 3C 454.3 as well (9.6\% vs. 6.4\% Acceptance Probability). While a best fit break in the PSD was measured at $\log(\nu_b) = -6.4$, Fig~\ref{fig_C} (left) clearly shows that this is simply where the Poisson noise begins to dominate as evidenced by the intersection of the dot-dashed line and the dashed line. Also the break was completely unconstrained with the 90\% confidence interval spanning the entire range of the parameter space. We also looked at the second peak in the Acceptance Probability which occurred at $\log(\nu_b) = -7.4$ and $\alpha = 0.2$ shown in the right column of Fig~\ref{fig_C}. The Acceptance Probability for this peak was 7.2\% so still an improvement from the unbroken model. The low frequency slope found for this fit would be much flatter than previously seen and could be analogous to the low frequency breaks seen in Galactic Black Holes (GBH) while in their low-hard state. However this break is only based on the 2 or 3 lowest frequency PSD points which are also the points that are the least statistically significant since they represent the longest timescales. Due to this and the low Acceptance Probability we conclude that there is no break in 3C 454.3.

3C 273 on the other hand resulted in a much better fit with the singly broken model with an acceptance probability of 68\% compared to 17\% for the unbroken model. The best fit break frequency and low frequency slope occurred at $\log(\nu_b) = -6.8$ and $\alpha = 0.6$ respectively. This corresponds to a timescale of 73 days or about 2.5 months. The break was well constrained between $\log(\nu_b)  = -6.5$ and -7.2 but the slope was only constrained to an upper limit of 1.3 which is still flatter than the slope found for the unbroken model at 1.35. 

For the rest of the sample, we also fit a singly broken model but only allowed $\log(\nu_b)$ to vary between -8.0 and -6.0 at 0.1 steps to determine an approximate lower limit for the break frequency and fixed $\alpha = 1.0$. The results of these fits are also displayed in Table~\ref{tbl_7}. 7 of the sources resulted in completely unconstrained break frequencies. Most of these sources were either the 4 sources we flagged as bad fits for the unbroken model, or had significantly flat slopes ($\alpha< 0.5$) so neither 1.0 nor 2.0 was a close approximate to the slope of the PSD. 

We estimate the lower limit of the break frequency on our sample to be $\log(\nu_b) = -6.35$. This was found by averaging the upper limits of the break frequencies for all of the sources where a lower limit was constrained including Cen A and NGC 4151.This corresponds to a timescale of 26 days and is very close to the upper edge of our sampled frequency band where $P_{noise}$ usually dominates.
\subsection{Energy Dependence of PSD}

\begin{deluxetable*}{lccccccccccc}
\tablecolumns{12}
\tabletypesize{\scriptsize}
\tablecaption{Best Fit Parameters: Energy Dependence \label{tbl_4}}
\tablewidth{0pt}
\tablehead{
\colhead{} & \multicolumn{3}{c}{$14-24$ keV} & \colhead{} & \multicolumn{3}{c}{$24-50$ keV} & \colhead{} & \multicolumn{3}{c}{$50-150$ keV} \\
\cline{2-4} \cline{6-8} \cline{10-12} \\
\colhead{} & \colhead{} & \colhead{$\log A_{0}$} & \colhead{} & \colhead{} &\colhead{} &\colhead{$\log A_{0}$} & \colhead{} & \colhead{} & \colhead{} & \colhead{$\log A_{0}$} & \colhead{} \\
\colhead{Name} & \colhead{$\alpha$} & \colhead{(Hz$^{-1}$)} & \colhead{$1-R$} & \colhead{} &  \colhead{$\alpha$} & \colhead{(Hz$^{-1}$)} & \colhead{$1-R$}  & \colhead{} &  \colhead{$\alpha$} & \colhead{(Hz$^{-1}$)} & \colhead{$1-R$} \\
\colhead{(1)} & \colhead{(2)} & \colhead{(3)} & \colhead{(4)} & \colhead{} & \colhead{(5)} & \colhead{(6)} & \colhead{(7)} & \colhead{} & \colhead{(8)} & \colhead{(9)} & \colhead{(10)}}
\startdata
Cen A & 1.15$^{+0.35}_{-0.5}$ & 4.17$\pm$0.04 & 0.317 & & 1.55$^{+0.35}_{-0.25}$ & 3.54$\pm$0.08 & 0.399 & & 1.55$^{+0.65}_{-0.35}$ & 3.66$\pm$0.08 & 0.561 \\
NGC 4151 & 1.10$^{+0.25}_{-0.3}$ & 4.35$\pm$0.04 & 0.577 & & 1.30$^{+0.3}_{-0.3}$ & 3.99$\pm$0.05 & 0.365 & & 1.05$^{+0.4}_{-0.3}$ & 4.27$\pm$0.05 & 0.817 \\
Mrk 421 & 0.95$^{+0.25}_{-0.25}$ & 5.43$\pm$0.04 & 0.943 & & 0.80$^{+0.3}_{-0.25}$ & 5.64$\pm$0.04 & 0.768 & & 0.70$^{+0.4}_{-0.35}$ & 5.90$\pm$0.04 & 0.977 \\
NGC 3227 & 0.65$^{+0.35}_{-0.4}$ & 5.47$\pm$0.05 & 0.735 & & 0.80$^{+0.55}_{-0.3}$ & 5.18$\pm$0.04 & 0.306 & & 0.70$^{+0.5}_{-0.3}$ & 6.00$\pm$0.04 & 0.006 \\
NGC 2110 & 0.55$^{+0.3}_{-0.3}$ & 5.00$\pm$0.05 & 0.675 & & 1.15$^{+0.4}_{-0.45}$ & 4.46$\pm$0.04 & 0.139 & & 0.70$^{+0.35}_{-0.2}$ & 4.69$\pm$0.04 & 0.674 \\
\enddata
\tablecomments{Best fitting parameters for the PSDs in each energy band.}
\end{deluxetable*}

\begin{figure*}[t!]
\centering
\includegraphics[width = 0.8\columnwidth]{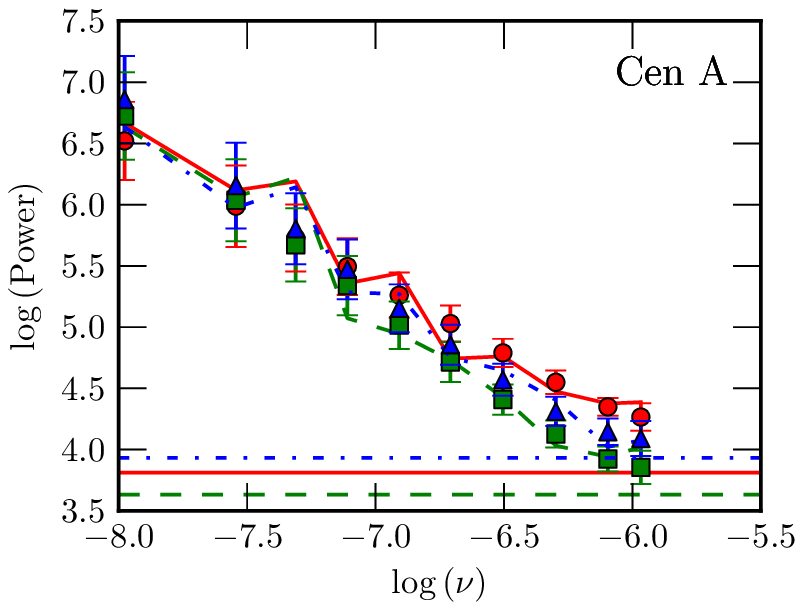}
\includegraphics[width = 0.8\columnwidth]{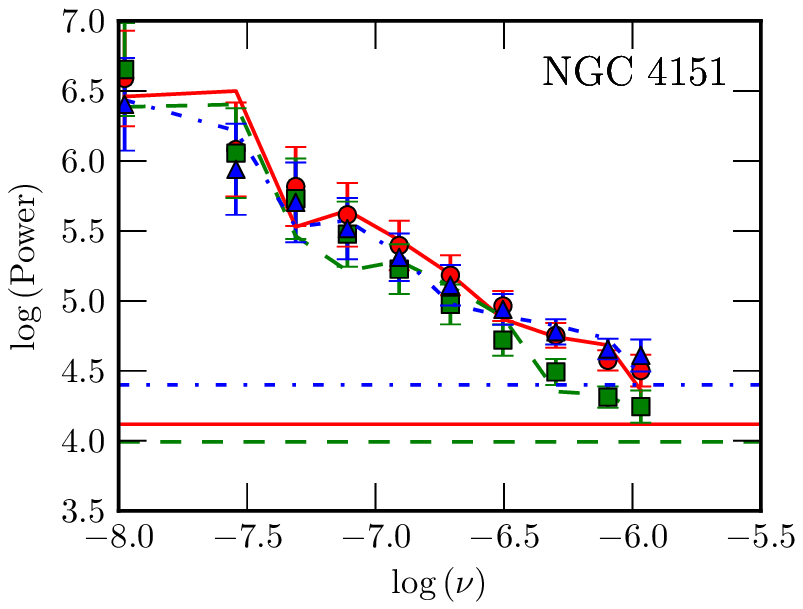}
\includegraphics[width = 0.8\columnwidth]{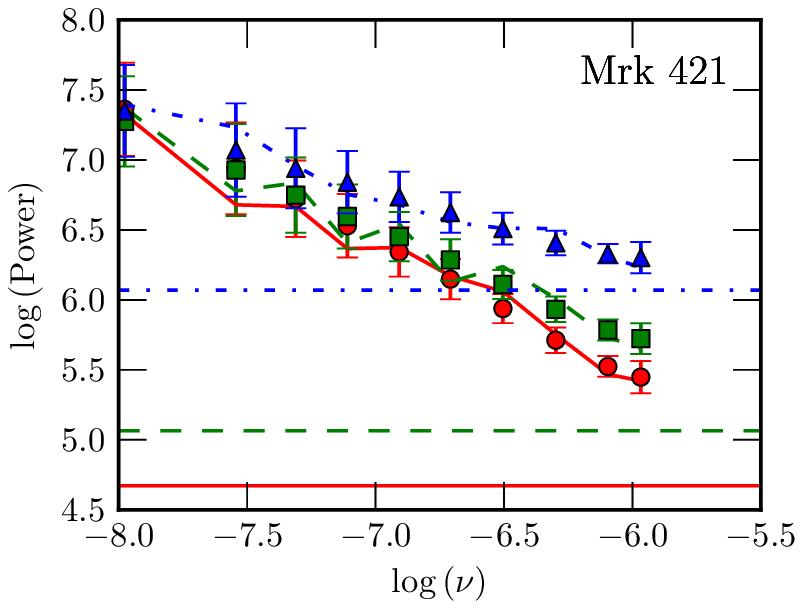}
\includegraphics[width = 0.8\columnwidth]{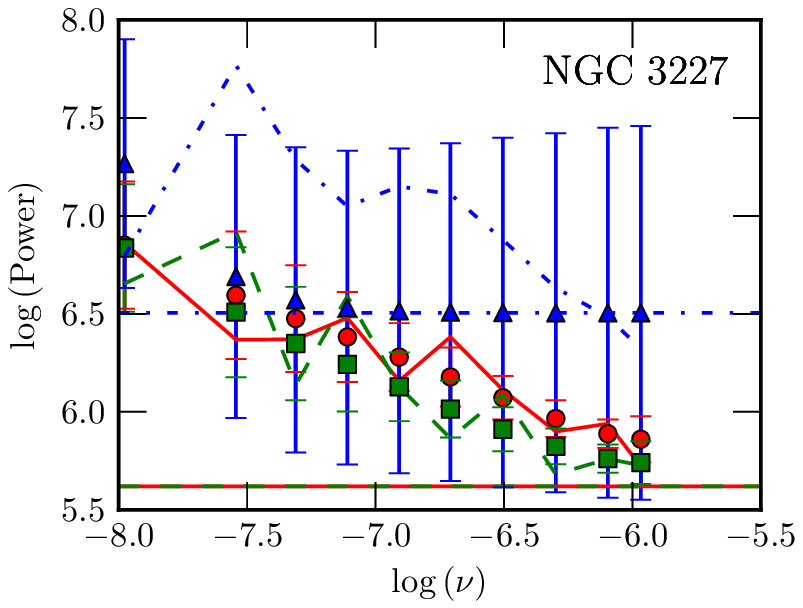}
\includegraphics[width = 0.8\columnwidth]{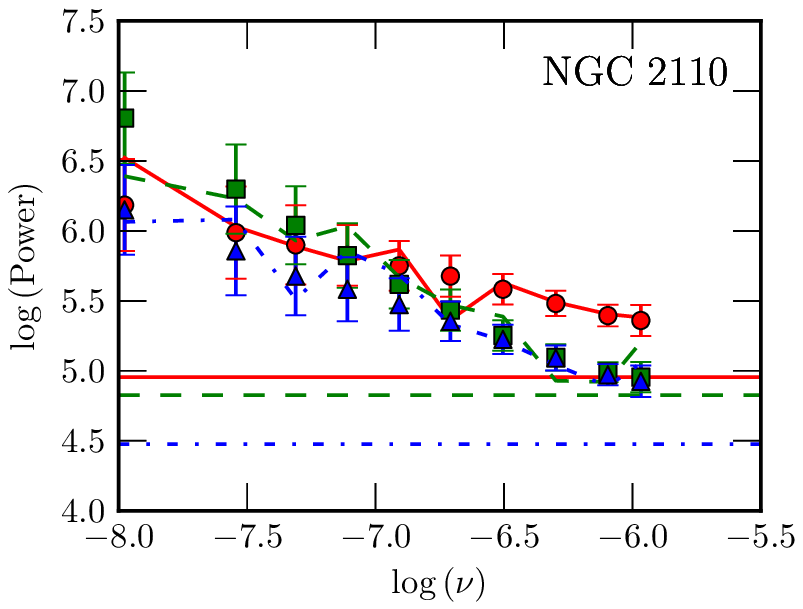}
\caption{Results from fitting the 14--24 keV, 24--50 keV, and 50--150 keV PSDs for Cen A,  NGC 4151, Mrk 421, NGC 3227, NGC 2110. Circles and solid lines correspond to the 14--24 band (colored red in the online version), squares and dashed lines correspond to the 24--50 keV band (colored blue in the online version), and the triangles and dot-dashed lines correspond to the 50--150 keV band. The non horizontal lines plot the observed PSD, the straight horizontal ones the Poisson noise level, and the markers with error bars are the best fit model PSD.\label{fig_18}}
\end{figure*}

For 5 sources (Cen A, NGC 4151, Mrk 421, NGC 3227, and NGC 2110), the S/N was high enough to split into three separate energy bands and determine best fit parameters for the PSD of each energy band. Table~\ref{tbl_4} lists the best fitting power law slope, normalization and Acceptance Probability for each source in each band. We split the 14-150 keV full band into 14-24 keV (Band 1), 24-50 keV (Band 2), and 50-150 keV (Band 3) bands. These represented the best bands to get nearly equal S/N in each band but also allows for possible separation of spectral components because reflection should approximately be most important in the 24-50 keV band. Figure~\ref{fig_18} plots the best fit models for each source in each band.

All of the AGN were fit very well with an unbroken power law again, except for NGC 3227 in the 50-150 keV band but this was due to poor S/N in the PSD rather than an ill-fitting model. The average power law slopes for each band are 0.88$\pm0.24$, 1.12$\pm0.29$, and 0.94$\pm0.33$ for  Band 1, 2, and 3 respectively with errors the standard deviation of the sample. There seems to be some evidence for the increase of the slope between bands 1 and 2, however it is only significant at the 1$\sigma$ level. Moreover, all three bands are consistent with a power law slope of 1.0 so we conclude there is no detectable dependence of the PSD with energy. 

\subsection{Correlation with BH Mass, X-ray Luminosity, Accretion Rate, and Column Density}\label{sub_corr}

\begin{deluxetable}{lcc}[t!]
\tablecaption{Correlation between Hard X-Ray Variability and Source Properties\label{tbl_5}}
\tablewidth{0pt}
\tablehead{\colhead{Variable} & \colhead{$\rho_s$} & \colhead{$P_{\rho}$}\\
\colhead{(1)} & \colhead{(2)} & \colhead{(3)}}
\startdata
\textbf{All Sample} & & \\
$L_{Bol}$ & 0.14 & 0.49 \\
$M_{BH}$ & 0.08 & 0.70\\
$L_{Bol}/L_{Edd}$ & -0.10 & 0.63 \\
$N_{H}$ & -0.16 & 0.44 \\
\textbf{Seyferts Only} & & \\
$L_{Bol}$ & -0.01 & 0.97 \\
$M_{BH}$ & -0.04 & 0.87 \\
$L_{Bol}/L_{Edd}$ & -0.23 & 0.32 \\
$N_{H}$ & -0.09 & 0.71
\enddata
\tablecomments{Col. (1): Source property used in correlation analysis with hard X-ray variance calculated by integrating the best fit PSD. Col. (2): $\rho_s$ is the Spearman rank correlation coefficient  Col. (3): $P_{\rho}$ the null probability.}
\end{deluxetable}

To determine any significant correlations with properties of the AGN, we used the integrated PSD as the dependent variable rather than any of the parameters of the PSD itself. Integrating over the best-fit PSD gives us an estimate of the excess variance produced by the source itself, in other words variance above that expected by measurement error and noise. Excess variance is then defined as

\begin{equation}\label{eq_10}
\sigma_{\rm XS}^{2} = \int_{\nu_{1}}^{\nu_{2}} P(\nu)d\nu
\end{equation}

\noindent which can be analytically integrated with an unbroken power law model of the form in Equation~\ref{eq_7} to find

\begin{equation}\label{eq_11}
\sigma_{\rm XS}^{2} = \frac{A_0}{(1 - \alpha)\nu_0^{-\alpha}}\left[\nu_2^{(1-\alpha)} - \nu_1^{(1-\alpha)}\right]
\end{equation}

\begin{figure}[t!]
\epsscale{1.0}
\plotone{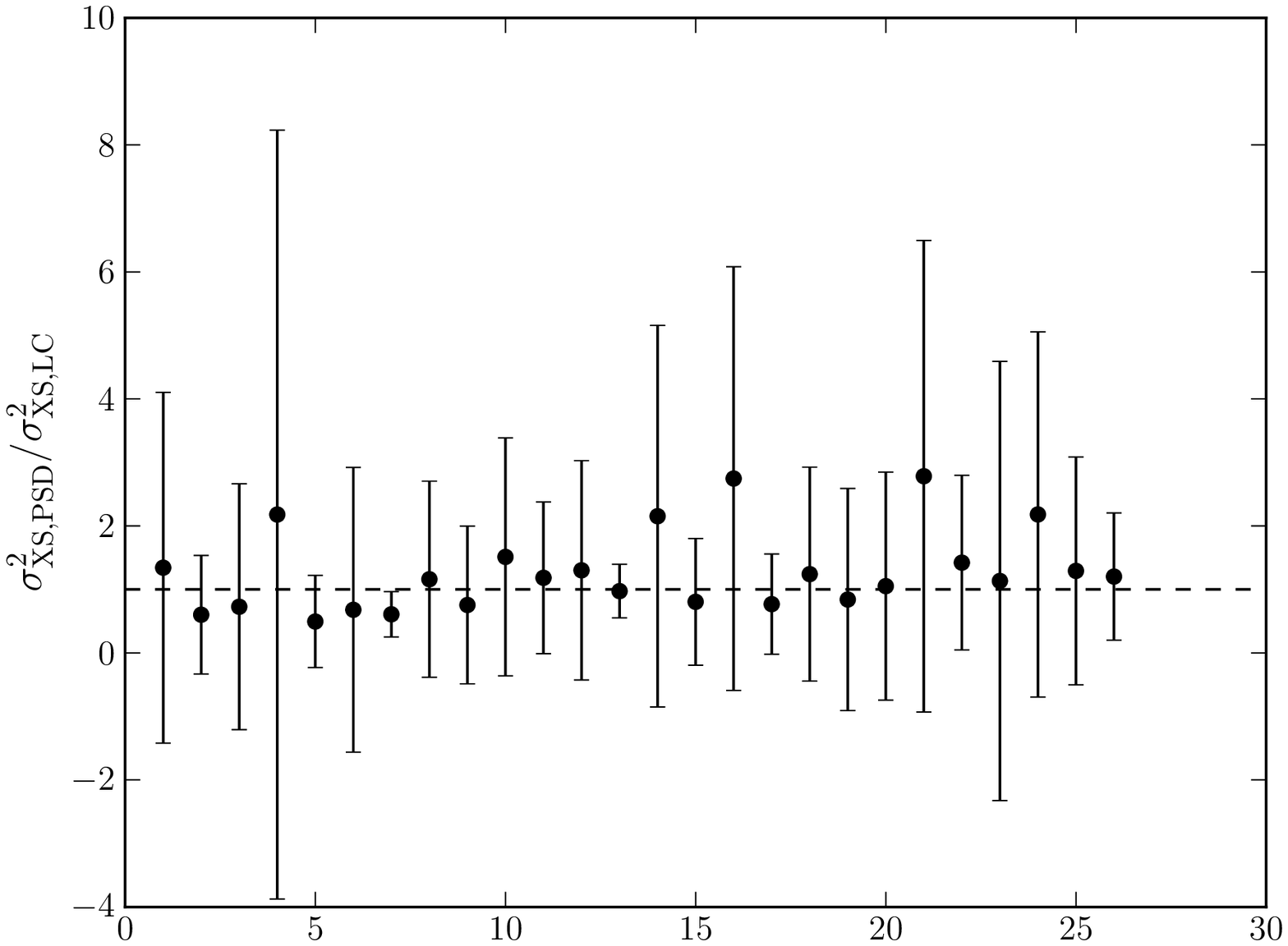}
\caption{Ratio of the excess variance determined by integrating the best-fit PSD and by using the variance of the light curves.  Dashed line indicates a ratio of 1.\label{fig_10}}
\end{figure}

For all of the AGN we substituted in for $A_0$ and $\alpha$ the best fit normalization and power law slope found in Table~\ref{tbl_3} however we excluded from the analysis the 4 AGN discussed previously that had bad fits to the PSD. To test that we were getting reasonable values of the excess variance, we determined it using the time domain definition \citep{1997ApJ...476...70N, 2002ApJ...568..610E}.

\begin{equation}\label{eq_12}
\sigma^2_{\rm XS} = \frac{S^2 - \overline{\sigma^2_{\rm err}}}{\mu^2}
\end{equation}

\noindent where $S^2$ is the sample variance

\begin{equation}\label{eq_13}
S^2 = \frac{1}{N-1}\sum^{N}_{i = 1}(x_i - \mu)^2
\end{equation}

and $\overline{\sigma^2_{\rm err}}$ and $\mu$ are defined as before. The errors on the integrated PSD variance were determined by propagating the errors on the normalization and the slope through Equation~\ref{eq_11} while also taking into account the anti-correlation between normalization and power law slope. The errors on the time domain variance were calculated using the following equation

 \begin{equation}\label{eq_14}
err(\sigma^2_{\rm XS}) = \sqrt{\left(\sqrt{\frac{2}{N}}\cdot\frac{\overline{\sigma^2_{\rm err}}}{\mu^2}\right)^2 + \left(\sqrt{\frac{\overline{\sigma^2_{\rm err}}}{N}}\cdot\frac{2F_{\rm var}}{\mu}\right)^2}
\end{equation}

where N is the number of light curve points, and $F_{\rm var}$ is the square root of $\sigma^2_{\rm XS}$ \citep{2003MNRAS.345.1271V}. But this only takes into account the error due to measurement uncertainties. There is also uncertainty due to the random scatter intrinsic to the red noise process that produces the variability. For this uncertainty we used Table 1 from \citet{2003MNRAS.345.1271V}. Our light curves contained 400 points and the PSDs had a slope of about 1, so we used an uncertainty of 0.18 which is halfway between the uncertainty of a light curve with 100 points and one with 1000 points \citep{2003MNRAS.345.1271V}. We then summed in quadrature the uncertainty from Equation~\ref{eq_14} and 0.18 to determine the total uncertainty. Figure~\ref{fig_10} shows the relationship between the PSD integrated and time domain $\sigma_{\rm XS}^2$. While there is some scatter in the relationship, it seems that both indicators agree fairly well with each other within the error.

\begin{figure*}[t!]
\centering
\includegraphics[width= 0.7\textwidth]{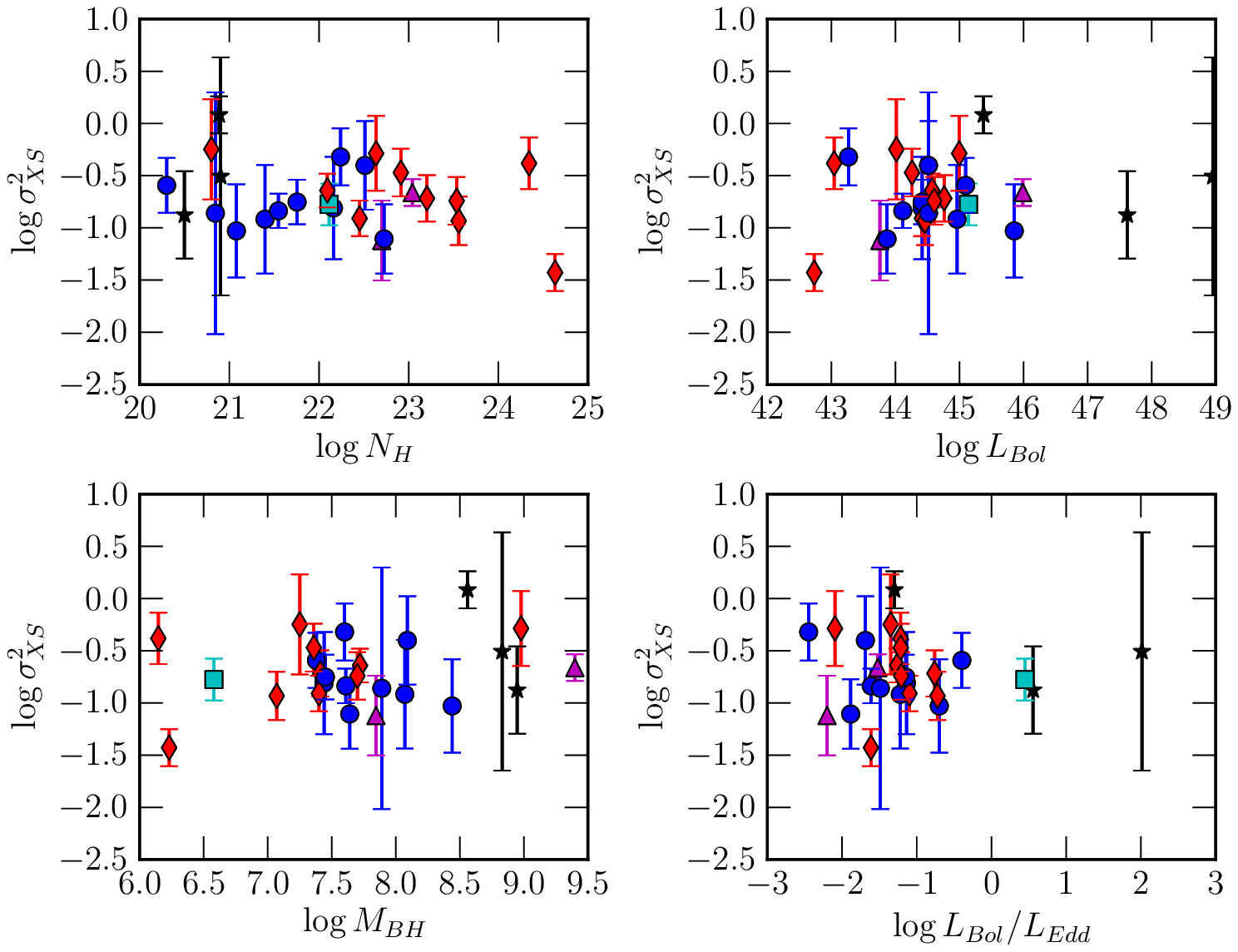}
\caption{Correlation between excess variance of the whole sample BAT AGN and their $N_{H}$ (top left), $L_{Bol}$ (top right), $M_{BH}$ (bottom left), and $L_{Bol}/L_{Edd}$ (bottom right). Circles (colored blue in online version) are Seyfert 1-1.5, diamonds (colored red in online version) are Seyfert 1.8-2, stars (colored black in online version) are Blazars, triangles (colored cyan in online version) are BLRG, and squares (colored magenta in online version) are NLRG. See Table~\ref{tbl_5} for results of statistical analysis. \label{fig_11}}
\end{figure*}

Table~\ref{tbl_5} shows the results of a Spearman rank correlation test between the hard X-ray variance and bolometric luminosity, black hole mass, accretion rate with $L_{\rm bol}/L_{\rm edd}$ as the indicator, and column density. We looked for correlations using both the whole sample of AGN and a sample of just Seyfert galaxies because radio-loud AGN could have variability introduced due to other processes associated with the jet especially in the energies studied here. Figures~\ref{fig_10} \&~\ref{fig_11} display the 4 correlations for both samples.

\begin{figure*}[t!]
\centering
\includegraphics[width = 0.7\textwidth]{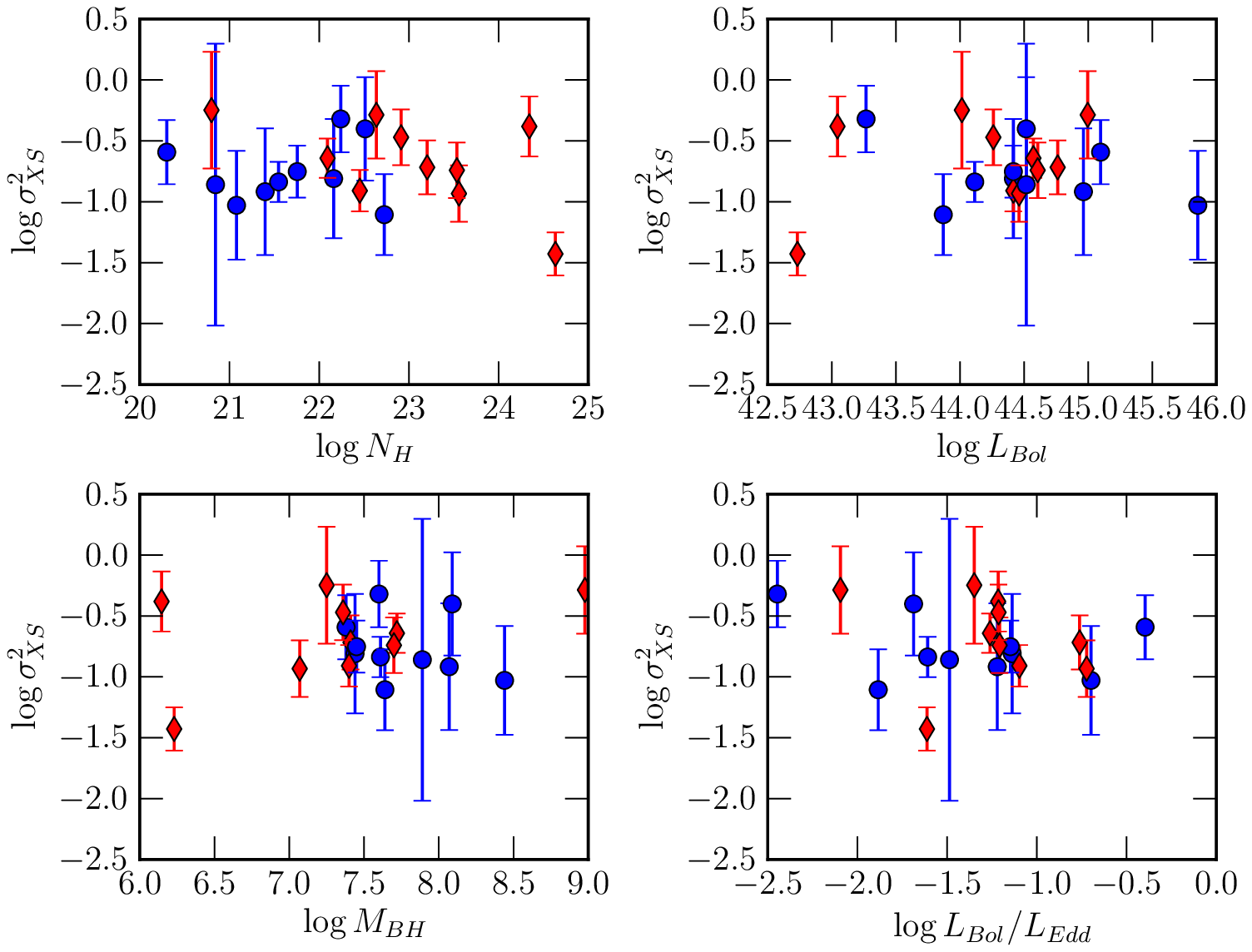}
\caption{Same as Figure~\ref{fig_11} except only using Seyfert galaxies. \label{fig_12}}
\end{figure*}

As can be seen in the Figures and Table, there seems to be no significant correlation between any of the source properties and hard X-ray variance. None of the correlations have a null hypothesis probability, $P_{\rho} <  0.3$ and the largest correlation coefficient is merely 0.23. This result agrees with those of \citet{2004ApJ...617..939M} and \citet{2007A&A...473..105M} that found the anticorrelation between variability amplitude and luminosity and black hole mass weakens as longer timescales are sampled. This can be explained as a consequence of the simple scaling of PSD break timescales with luminosity and black hole mass. On short timescales, the difference in PSD breaks between different AGN will manifest itself when determining variability amplitude since this is merely an integration of the PSD. However as one moves to longer and longer timescales away from the PSD break, variability amplitude increases (as it should for red noise processes) but eventually saturates, and the dependence on luminosity and black hole mass disappears. This makes sense in our sample as well since we report only one object where the PSD break is in the frequency band sampled by the BAT light curves.

\section{Discussion}\label{sect_5}
\subsection{Comparison with 2--10 keV Variability}

\begin{deluxetable*}{lccccccccc}
\tablecolumns{9}
\tabletypesize{\scriptsize}
\tablecaption{2--10 keV PSD Parameters\label{tbl_6}}
\tablewidth{0pt}
\tablehead{\colhead{} & \multicolumn{5}{c}{2--10 keV} & \colhead{} & \multicolumn{3}{c}{14--150 keV}\\
\cline{2-6} \cline{8-10}
\colhead{Name} & \colhead{$\alpha_{2-10}$} & \colhead{$\log \nu_b$} & \colhead{$\beta_{2-10}$} & \colhead{$A_1$} & \colhead{Reference} & \colhead{} & \colhead{$\alpha_{14-150}$} & \colhead{$A_0$ or $A_1$} & \colhead{$\log \nu_b$}\\
\colhead{(1)} & \colhead{(2)} & \colhead{(3)} & \colhead{(4)} & \colhead{(5)} & \colhead{(6)} & \colhead{} & \colhead{(7)} & \colhead{(8)} & \colhead{(9)}}
\startdata
3C 273 & 1.0 & -6.2 & 2.4 & $10^4$ & M04 & & 0.6$^{+0.7}_{-\infty}$ & $3.16\times10^5$ & $-6.8^{+0.3}_{-0.4}$\\
3C 390.3 & 1.0 & $-6.6^{+0.4}_{-0.2}$ & $2.4^{+0.3}_{-0.3}$ & $2.51\times10^4$ & G09 & & $0.75^{+0.65}_{-0.55}$ & $3.42\times10^4$  & \nodata\\
Cen A & $0.9^{+0.3}_{-0.2}$ & $-6.2^{+0.3}_{-0.2}$ & $2.5^{+0.4}_{-0.1}$ & $7.90\times10^4$ & R11 & & $1.35^{+0.35}_{-0.35}$ & $6.61\times10^3$ & \nodata\\
Circinus Galaxy & 1.0 & $-3.39^{+0.13}_{-0.13}$ & $3.23^{+0.12}_{-0.12}$ & 8.10 & GM12 & & $0.4^{+0.35}_{-0.4}$ & $2.39\times10^{4}$ & \nodata\\
NGC 3227 & 1.0$^{+0.2}_{-0.4}$ & $-4.59^{+0.36}_{-0.158}$ & $>1.9$ & 371 & U05 & & $0.80^{+0.4}_{-0.35}$ & $1.59\times10^5$ & \nodata\\
NGC 3516 & $1.1^{+0.4}_{-0.3}$ & $-5.7^{+0.4}_{-0.3}$ & $2.0^{+0.55}_{-0.2}$ & $7.90\times10^3$ & Ma03 & & $0.35^{+0.4}_{-0.35}$ & $9.96\times10^4$ & \nodata\\
NGC 3783 & $0.80^{+0.8}_{-0.5}$ & $-5.2^{+0.88}_{-1.01}$ & $2.6^{+0.6}_{-1.0}$ & $1.10\times10^3$ & S07 & & $0.5^{+0.3}_{-0.5}$ & $9.83\times10^4$ & \nodata\\
NGC 4151 & $1.1^{+1.35}_{-0.4}$ & $-5.9^{+0.4}_{-0.7}$ & $2.1^{+1.9}_{-0.25}$ & $1.30\times10^4$ & Ma03 & & $1.1^{+0.45}_{-0.25}$ & $1.34\times10^4$& \nodata\\
NGC 5548 & $1.15^{+0.5}_{-0.65}$ & $-6.22^{+0.60}_{-0.40}$ & $2.05^{+0.8}_{-0.4}$ & $2.50\times10^4$ & Ma03 & & $0.95^{+2.05}_{-0.45}$ & 1.01$\times10^5$ & \nodata
\enddata
\tablecomments{Comparison between the 2--10 keV PSDs found from a literature search and 14--150 keV PSDs from this work. Col. (1): Name of source. Col. (2): Low frequency power law slope for 2--10 keV PSD. Col. (3): Break frequency for 2--10 keV PSD. in Hz. Col. (4): High frequency power law slope for 2--10 keV PSD. Col. (5): Normalization for 2--10 keV PSD in Hz$^{-1}$ Note that the normalization here is defined as the variability power at the break frequency, not at $10^{-6}$ Hz like in this work. Col. (6): Reference for the 2--10 keV PSD. Col. (7) and (8): 14--150 keV PSD parameters found in this work.}
\tablerefs{M04: \citet{2004AIPC..714..167M}; G09: \citet{2009ApJ...703.1021G}; R11: \citet{2011ApJ...733...23R}; GM12: \citet{2012A&A...544A..80G}; U05: \citet{2005MNRAS.363..586U}; Ma03: \citet{2003ApJ...598..935M}; S07: \citet{2007MNRAS.378..649S}}
\end{deluxetable*}

\begin{figure*}[t!]
\centering
\includegraphics[width = \textwidth]{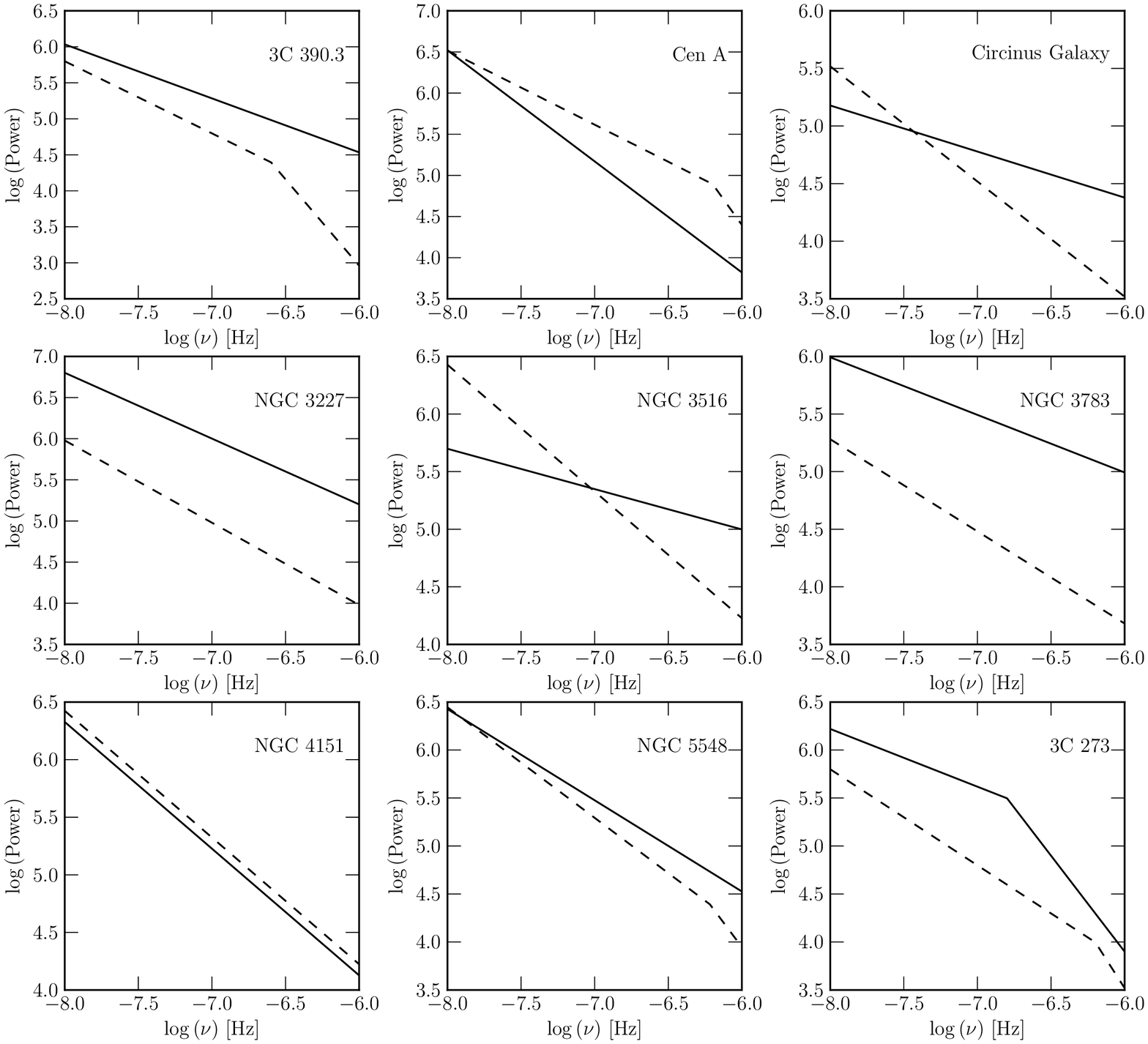}
\caption{Comparison between the PSDs measured in the 2--10 keV band (dashed) and the 14--150 keV band (solid).\label{psd_comp}}
\end{figure*}

\begin{figure}[t!]
\epsscale{1.0}
\plotone{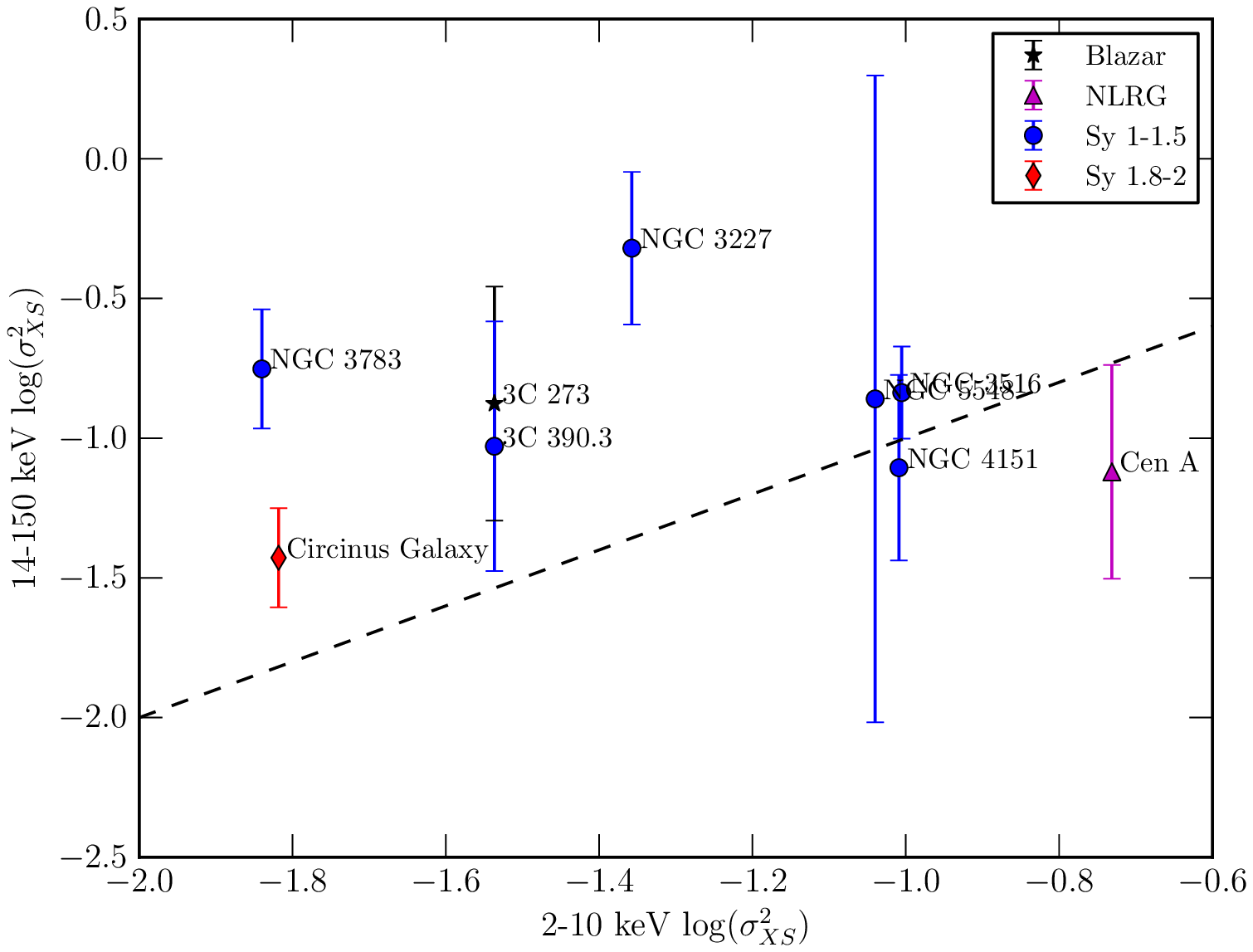}
\caption{Comparison between the 2--10 keV excess variance and 14--150 keV excess variance. All of the sources have larger or nearly equal variability power in the higher energy compared to lower energy. \label{fig_13}}
\end{figure}

We cross-referenced our sample with the samples of previous PSD studies in the 2--10 keV band and found 9 AGN's PSD had previously been measured. Table~\ref{tbl_6} lists these AGN and their PSD parameters. Most of the 2--10 keV studies used {\it RXTE} light curves to determine low frequency power and either XMM or Chandra light curves for the high frequency power. \citet{2012A&A...544A..80G} (GM12) used shorter length XMM light curves so the observed PSD only covered the high frequencies, and when the authors fit the PSD, they fixed the low frequency slope to a value of -1.0. Figure~\ref{psd_comp} shows the PSDs measured in the two bands for the timescales sampled in this study. 

In general, there does not seem to be a significant difference in the power law slopes of the soft and hard X-ray PSDs. We find an average slope of 0.8, similar to the low frequency slope of $\sim$1 found in the lower energy studies \citep{2002MNRAS.332..231U,2003ApJ...593...96M,2004MNRAS.348..783M,2005MNRAS.363..586U,2007ApJ...656..116M,2009ApJ...698.1740M}. \citet{2011ApJ...730...52K}, using Bayesian analysis and modeling light curves as a linear combination of Gaussian Ornstein--Uhlenbeck processes determined the slopes and break frequencies of optical and soft X-ray light curves of AGN. This modeling technique allows for a better interpretation of the actual physical processes occurring in the AGN. They also find a power law slope of 1.0 describes very well the long timescale PSD in the 2--10 keV energy range. This suggests the long timescale variability process is the same between the 2--10 and 14--150 keV energy bands.

We also compared the intrinsic variances of the AGN in both energy bands, again using the excess variance calculated directly from the light curves to get a better estimate of the uncertainty. For the 2--10 keV PSDs, we integrated the PSDs we found in the literature, taking into account the break frequency when needed. Some of the AGN were best fit with a slowly bending power law model instead of a sharp break, however when integrating the PSDs, we strictly used the sharp break model to get a fairly close estimate, especially since the break is usually near the edge of our frequency range or outside it completely. 

Figure~\ref{fig_13} shows a comparison between the variances of the two energy bands with a dashed line indicating equal variances. All but two sources (NGC 4151 and Cen A) show an apparent increase in variance when the energy band increases from 2--10 keV to 14--150 keV. This result agrees with that found by \citet{2012A&A...537A..87C} which only looked at the brightest AGN in the BAT sample and is surprising because variance had been seen previously to decrease as energy increases \citep{2004MNRAS.348..783M,2011PASJ...63S.669I}. Using {\it Suzaku} light curves, \citet{2009PASJ...61.1331M} studied the temporal and spectral variations of 36 AGN in 3 different energy bands (0.5--2 keV, 2--10 keV, and 15--50 keV). Their sample overlaps with our sample, and their highest energy band overlaps the BAT band. They as well found that the 15--50 keV variability is much lower than that in the lower energies. However all of these studies used much shorter timescales than the one presented here. Whereas our timescales are on the order of years, previous studies are only on scales of less than half a day. This seems to indicate that the variability process is timescale and energy dependent, where at shorter timescales lower energies display higher variability and on longer timescales variability is at least equal across all energies. 

For 3C 273 we were able to compare the break frequencies between the 2--10 keV and 14--150 keV band. Interestingly the break frequency seems to have decreased from $\log \nu_b$ = -6.2 to -6.8 (18 days to 73 days).  However as shown in Figure~\ref{fig_13} the total variance possibly increased or at the least stayed the same. This is probably due to the decrease in both the low and high frequency slope where the low frequency slope decreased from 1.0 to 0.6 (although only an upper limit of 1.3 could be constrained) and the high frequency slope decreased from 2.4 to 2.0 (although this is because we fixed the slope at 2.0). We caution the reader that without error bars from the 2--10 keV study, a valid comparison cannot be made.

\subsection{Model for Variability?}\label{model_sec}
\cite{1997MNRAS.292..679L} suggests a model for the variability of accreting objects in which inwardly propagating fluctuations of the accretion rate couple at each radii of the accretion disk and modulate the X-ray emission. This model successfully explains the linear rms-flux relation seen in all GBH's and AGN \citep{2001MNRAS.323L..26U} and can explain the energy dependent time lags \citep{2001MNRAS.327..799K}. \citet{2006MNRAS.367..801A} further investigate this model and determine that if the emissivity of the X-ray emitting corona is radially dependent, then there should be a change in the shape of the observed PSD with energy. The break frequency correlates with the radius at which the emissivity drops off, thus for example, if the harder X-rays originate from a more compact region than softer X-rays, then the break frequency should increase as energy increases since there will be more variability on shorter timescales due to the small radius. 

Our results do not rule out the possibility of this model but only due to the fact that we detected only one break frequency in 3C 273. The BAT PSDs were limited to the longest of timescales with the shortest timescale being $\sim$week. This is about the longest timescale that a break has been measured in softer bands with most breaks occurring on shorter timescales \citep{2012A&A...544A..80G}. The fact that the power law slope stays roughly constant around 1.0 does agree with the results of \citet{2006MNRAS.367..801A} who found a constant low frequency slope with energy.

3C 273's measured break frequency, while seeming to decrease with energy, is still formally consistent between the 2--10 and 14--150 keV band especially given the lack of error bars from \citet{2004AIPC..714..167M}. Because of the large uncertainty though, the possibility remains that 3C 273 exhibits different variability mechanisms in the two bands, maybe due to its radio jets.

\citet{2009A&A...504...61I} recently proposed another model for both accretion onto a black hole and variability. Instead of a smooth accretion disk flowing into the black hole, they use clumps of material falling in. These clumps collide and first form optically thick shocks that produce the optical/UV photons typically seen in AGN and GBH's. The optically thick shocks then become optically thin while the whole time electrons are being heated. The optical/UV photons originally emitted in the optically thick shock provide the seed photons for inverse Comptonization and are upscattered into X-rays much the same way in the usual accretion scenario. Variability however is attributed to the actual Comptonization process, and the PSD break timescale is associated with the Compton cooling timescale of the heated electrons. This model has been shown to reproduce the empirically found correlations between PSD timescale, black hole mass, and accretion rate through simple physics \citep{2012A&A...540L...2I}. 

However our results with their limited frequency range are not able to discern between the two different models. What is needed are high quality light curves at these high energies on shorter timescales to be able to measure break frequencies in a large sample of AGN.

\subsection{Variance vs. Type}

\begin{figure}[t!]
\epsscale{1.0}
\plotone{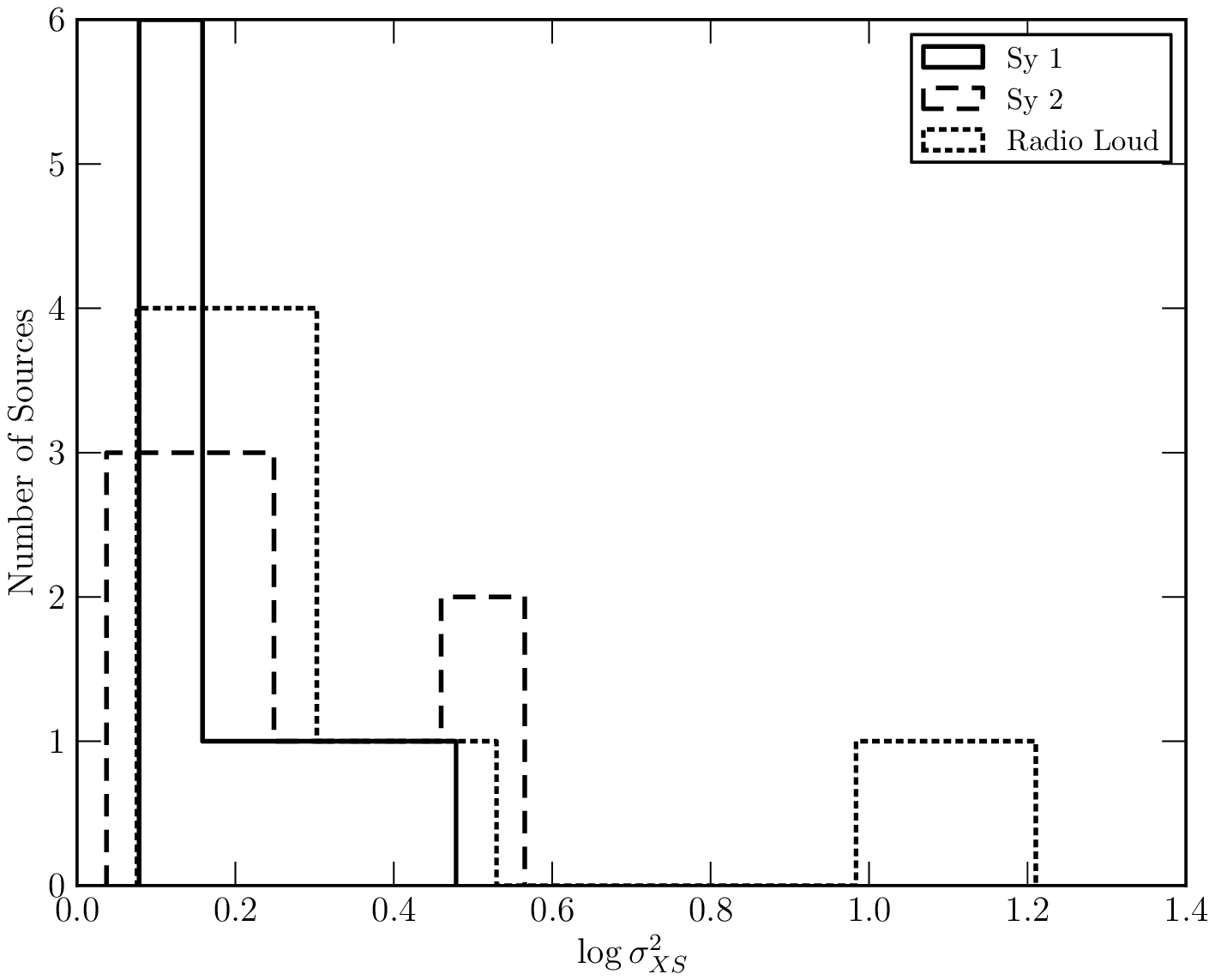}
\caption{Distribution of the excess variance by type of source: Seyfert 1, Seyfert 2, and radio loud AGN. We grouped all Seyfert 1, 1.2, and 1.5 as Seyfert 1s and Seyfert 1.8, 1.9, and 2 and Seyfert 2s. Radio loud objects include Blazars, BLRG, NLRG.\label{fig_14}}
\end{figure}

Figure~\ref{fig_14} plots the histograms of excess variances based on the type of source. As before we grouped all Seyfert 1, 1.2, and 1.5 as Seyfert 1 and all Seyfert 1.8, 1.9, and 2 as Seyfert 2. But we also grouped all Blazars, NLRG, and BLRG as radio loud objects. From the histograms there does not appear to be any difference in the variances of the different types of AGN. The one object with the largest variance is Mrk 421 which has been known to exhibit extremely strong variability in the form of flares over many timescales \citep{2004ApJ...605..662C}. However it is interesting to note that even though Mrk 421 is the only source where radiation is primarily beamed,  the slope of the PSD for Mrk 421 (0.8) is not different than the slopes of any of the other AGN,  it is just the normalization that is significantly higher. This suggests that long timescale variability might be caused by the same process no matter the type of AGN which agrees with the results seen in \S~\ref{sub_corr}.

\subsection{Comparison with GBH}
GBH's due to their closeness and large fluxes have much higher quality spectra. They exhibit state transitions characterized by their intensity (high or low) and shape (soft or hard) of the energy spectrum. The different states also show different variability that change the PSD \citep{1997ApJ...474L..57C, 2003A&A...407.1039P}. In the high/soft state, the PSD is best fitted by a singly-broken or bending power law while in the low/hard state it is best fitted by a doubly-broken power law. For nearly all AGN, in the 2--10 keV band, their PSD are similar to the high/soft state of GBH because they are best described by a singly broken power law (see  \citet{2004MNRAS.348..783M} and \citet{2003ApJ...593...96M}). Currently the only AGN found to be best fit by a doubly-broken power law is Ark 564 \citep{2001ApJ...550L..15P,2002A&A...382L...1P,2007MNRAS.382..985M}.

Since our PSD's don't show any evidence for a low frequency break, they are consistent with the PSD's of GBHs in the high/soft state if the PSD is constant with energy. This is in agreement with studies in the 2--10 keV band including \citet{2003ApJ...593...96M} and \citet{2005MNRAS.363..586U}. Previous work looking at the energy dependence of Cygnus X-1's high state PSD also observed a constant low frequency power law slope with a possible increasing break frequency with energy \citep{1999ApJ...510..874N,2004MNRAS.348..783M}. Our finding that the PSD slope is relatively constant with energy agrees with this, and although we did not detect any break frequencies, the lack of detection indicates a lower limit for the break frequency at 10$^{-6}$. With the exception of 3C 273, our study seems to indicate that all of our AGN are in the high state since we don't find any evidence for a low frequency break that distinguishes between the low and high state. \citet{2011ApJ...726...21Z} used archival {\it RXTE} All Sky Monitor light curves to determine the variance at roughly the same timescales as our BAT light curves. They also find that the measured variability seems to indicate that AGN are scaled up versions of only the high states of GBHs. They determine that if AGN were in the low state with a low frequency break timescale,  there would still be an observed correlation between the variance and black hole mass and luminosity whereas the high state with a singly broken PSD would not produce any correlation at long timescales. This is exactly what we observe in our sample at higher energies using the more rigorous PSD analysis.

\section{Conclusions}\label{sect_6}
We have for the first time calculated and fit the PSDs of AGN in the hard X-ray (14--150 keV) regime. The PSDs were fit very well with an unbroken power law model with an average slope of $\sim-1$ and only one object, 3C 273, was found to require a break in the PSD. The lack of break frequencies at the longest of timescales points toward a majority of AGN being in the high state seen in GBHs. The slopes found are similar to those found in 2--10 keV PSD studies and the total variance at these long timescales does not show any evidence of changing with energy.

We did not find any correlation between the hard X-ray variance and properties of AGN previously seen in lower energy variability studies. It seems that whatever process is producing long timescale variability is completely independent of the standard measures of AGN including luminosity, black hole mass, accretion rate, and column density. This also holds when looking at the distribution of the strength of variability with AGN classification.

Unfortunately our hard X-ray variability study has not provided much more insight into verifying the processes that are producing the variability since only one PSD break was detected. However we have provided an upper limit of $\sim26$ days for the break timescale of the hard X-ray PSDs leaving open the possibility that the break timescale still decreases with energy (break frequency increases). 

Further work needs to be done in extending the frequency range of the PSDs. With the recent launch of NuStar this will be possible because it will be much more sensitive than BAT. Using WebPimms we predict for a source with flux $2\times10^{-11}$ ergs/s/cm$^{2}$ a count rate of .07 cts/s compared to $3\times10^{-5}$ cts/s for BAT. This will allow us to sample the high temporal frequencies just as XMM provided for the 2--10 keV band. Combining the BAT and NuStar PSDs would give us a broadband PSD spanning at least 6 decades of frequency allowing for better constraints on models for AGN. It will also allow us to significantly constrain the break frequency if one exists and determine if the same relation between black hole mass, accretion rate, and break frequency exists in the hard X-ray band as does in the soft X-ray band found by \citet{2006Natur.444..730M}.

\acknowledgments
The authors thank the anonymous referee for their detailed comments and suggestions that improved the quality of the paper. TS thanks Alex Markowitz for his help in understanding and implementation of the PSRESP fitting algorithm and also helpful comments on the paper. TS would also like to thank Abdurahmen Zoghbi for his comments and discussion of the paper. The authors thank Jack Tueller and the BAT team for their hard work in processing and generating the BAT light curves. 


\end{document}